\documentclass{aa}  

\usepackage[varg]{txfonts}  
\usepackage{amsmath} 
\usepackage{graphicx}

\usepackage{version}
\excludeversion{confidential}

\usepackage[dvipsnames]{xcolor}
\usepackage{natbib}

\usepackage{graphicx}
\usepackage[labelfont=bf,font=small]{caption}
\usepackage{subcaption}
\usepackage{amsmath}
\usepackage{txfonts}

\usepackage{color}
\usepackage{colortbl} 
\usepackage[hyphenbreaks]{breakurl}

\newcommand{\Msun}{\ensuremath{\,M_\odot}\xspace}
\newcommand{\Rsun}{\ensuremath{\,R_\odot}\xspace}
\newcommand{\Lsun}{\ensuremath{\,L_\odot}\xspace}
\newcommand{\kms}{\ensuremath{\,\rm{km}\,\rm{s}^{-1}}\xspace}

\newcommand{\lang}[1]{{\color{black}{#1}}}   

\renewcommand\eqref[1]{\ifnum\ifhmode\spacefactor\else2000\fi>1000 Equation~\ref{#1}\else Eq.~\ref{#1}\fi}
\newcommand\eqreftwo[2]{\ifnum\ifhmode\spacefactor\else2000\fi>1000 Equations~\ref{#1} and \ref{#2}\else Eqs.~\ref{#1} and \ref{#2}\fi}

\newcommand\figref[1]{\ifnum\ifhmode\spacefactor\else2000\fi>1000 Figure~\ref{#1}\else Fig.~\ref{#1}\fi}
\newcommand\figreftwo[2]{\ifnum\ifhmode\spacefactor\else2000\fi>1000 Figures~\ref{#1} and \ref{#2}\else Figs.~\ref{#1} and \ref{#2}\fi}
\newcommand\figrefthree[3]{\ifnum\ifhmode\spacefactor\else2000\fi>1000 Figures~\ref{#1}, \ref{#2}, and \ref{#3}\else Figs.~\ref{#1}, \ref{#2}, and \ref{#3}\fi}

\newcommand\secref[1]{\ifnum\ifhmode\spacefactor\else2000\fi>1000 Section~\ref{#1}\else Sect.~\ref{#1}\fi}
\newcommand\secreftwo[2]{\ifnum\ifhmode\spacefactor\else2000\fi>1000 Sections~\ref{#1} and \ref{#2}\else Sects.~\ref{#1} and \ref{#2}\fi}
\newcommand\secrefthree[3]{\ifnum\ifhmode\spacefactor\else2000\fi>1000 Sections~\ref{#1}, \ref{#2}, and \ref{#3}\else Sects.~\ref{#1}, \ref{#2}, and \ref{#3}\fi}

\newcommand\appref[1]{\ifnum\ifhmode\spacefactor\else2000\fi>1000 Appendix~\ref{#1}\else Appendix~\ref{#1}\fi}

\newcommand\tabref[1]{\ifnum\ifhmode\spacefactor\else2000\fi>1000 Table~\ref{#1}\else Table~\ref{#1}\fi}

\usepackage{booktabs}    

\begin{document} 

\title{Ionizing spectra of stars that lose their envelope through interaction with a binary companion: Role of metallicity}
\titlerunning{Ionizing spectra of stars that lose their envelope through interaction with a binary companion}
   \author{Y.\ G\"{o}tberg\inst{1} \and S.\,E.~de Mink\inst{1} \and J.\,H.~Groh\inst{2} }
        \authorrunning{ G\"{o}tberg, De Mink \& Groh}  

   \institute{Anton Pannekoek Institute for Astronomy, University of Amsterdam, 1090 GE Amsterdam, The Netherlands \\
              \email{Y.L.L.Gotberg@uva.nl, S.E.deMink@uva.nl}
                       \and
   School of Physics, Trinity College Dublin, The University of Dublin, Dublin 2, Ireland
    \\
    \email{jose.groh@tcd.ie}}
   
   \date{Received ......; accepted ......}

\abstract
{
Understanding ionizing fluxes of stellar populations is crucial for various astrophysical problems including the epoch of reionization. Short-lived massive stars are generally considered as the main stellar sources. We examine the potential role of less massive stars that lose their envelope through interaction with a binary companion. Here, we focus on the role of metallicity ($Z$). For this purpose we \lang{used} the evolutionary code MESA and \lang{created} tailored atmosphere models with the radiative transfer code CMFGEN. 
We show that typical progenitors, with initial masses of 12\Msun, produce hot and compact stars ($\sim 4\Msun$, 60--80\,kK, $\sim$1\Rsun). These \lang{stripped stars} copiously produce ionizing photons, emitting 60--85\% and 30--60\% of their energy as HI and HeI ionizing radiation, for $Z=$0.0001--0.02, respectively. Their output is comparable to what massive stars emit during their Wolf-Rayet phase, if we account for their longer lifetimes and the favorable slope of the initial mass function. Their relative importance for reionization may be further favored since they emit their photons with a time delay ($\sim 20$\,Myrs after birth in our fiducial model). This allows time for the dispersal of the birth clouds, allowing the ionizing photons to escape into the intergalactic medium. 
At low $Z$, we find that Roche stripping fails to fully remove the H-rich envelope, \lang{because of} the reduced opacity in the subsurface layers. This is in sharp contrast with the assumption of complete stripping that is made in rapid population synthesis simulations, which are widely used to simulate the binary progenitors of supernovae and gravitational waves. Finally, we discuss the urgency to increase the observed sample of stripped stars to test these models and we discuss how our predictions can help to design efficient observational campaigns.
}


   \maketitle
%

\section{Introduction}\label{sec:introduction}

Massive stars have played many important roles since the earliest epochs of star formation. \lang{These stars} shape, heat\lang{,} and stir their surroundings \lang{and} play a key role in driving the evolution of their host galaxies \citep[e.g.][]{2014MNRAS.445..581H}. Over cosmic time, subsequent generations of massive stars chemically enriched the Universe with elements synthesized by nuclear fusion \citep[e.g.][]{2007PhR...442..269W}, slowly increasing the average metallicity (i.e.\ \lang{mass} mass fraction of elements heavier than H and He) of subsequent stellar populations. 

During their short lives, massive stars copiously produce (far) ultraviolet (UV) photons. Of particular interest are their photons with wavelengths shorter than the Lyman limit ($\lambda < 912$~\AA, i.e.\ with energies exceeding the ionization potential of hydrogen, $h\nu > 13.6$~eV). In the local Universe, massive stars are observed to ionize their immediate surroundings, giving rise to luminous HII regions \citep{2012flhs.book.....C}. At larger distances, they dominate the \lang{rest-frame} UV part of the integrated spectra of \lang{star-forming} galaxies and give rise to various strong emission lines \citep[e.g.][]{2001ApJ...556..121K}. Going back further in distance and time, the early generations of massive stars were the prime sources of ionizing photons emitted by the first galaxies \citep{2011ARA&A..49..373B}. These galaxies are held responsible for the large-scale phase transition, known as the \lang{Epoch of Reionization}, during which the intergalactic neutral hydrogen gas became ionized \citep{1997ApJ...483...21H,2012ApJ...752L...5B,2016ARA&A..54..313M}. 

Massive stars are frequently found in binary or multiple systems \citep[e.g.][]{1980ApJ...242.1063G, 2009AJ....137.3358M, 2014ApJS..215...15S}.  Recent studies have shown that, in the majority of cases, massive stars have a companion that is so close that severe interaction between the two stars at some point during their lives is inevitable \citep[e.g.][]{2007ApJ...670..747K,2012Sci...337..444S}. Similar conclusions have been reached by various groups using different datasets \citep{2012MNRAS.424.1925C, 2013A&A...550A.107S, 2014ApJS..213...34K, 2015A&A...580A..93D, 2017A&A...598A..84A}.

Binary evolution can lead to many complex evolutionary paths involving one or more phases of mass exchange between the two stars \citep[e.g.][]{1966AcA....16..231P, 1967ZA.....65..251K} and possibly a merger of the two stars. This has drastic consequences for the observable properties of both stars, their remaining lifetime and final fate. This raises the question about the implications for the integrated spectra of stellar populations, and their ionizing fluxes in particular. 

The most widely used spectral population synthesis codes, Starburst99 \citep{1999ApJS..123....3L, 2014ApJS..212...14L},  GALAXEV \citep{2003MNRAS.344.1000B} and FSPS \citep{2009ApJ...699..486C}, do not account for the possible effects of interacting binary stars and their products. In these simulations the ionizing \lang{flux comes} primarily from the most massive and luminous stars, which are short-lived. \lang{At} birth these stars \lang{already} emit a significant portion of their light at wavelengths shorter than the Lyman limit. The most massive and luminous single stars lose their hydrogen-rich envelope through stellar winds and eruptions. After a few million years they become Wolf-Rayet (WR) stars. During the WR phase, the stars have higher emission \lang{rates} of ionizing photons, but as the stars are still hot during the long-lasting main sequence, the total contribution from main sequence \lang{O-} and early B-type stars dominates the emitted ionizing photons from single stellar populations.

Pioneering efforts to account for the effects of binaries on the spectra of stellar populations have been undertaken by three major groups using the Brussels PNS code \citep{2003A&A...400...63V, 2007ApJ...662L.107V}, the Yunnan simulations \citep{2004A&A...415..117Z,2010Ap&SS.329...41H, 2010Ap&SS.329..277C, 2012MNRAS.421..743Z, 2012MNRAS.424..874L, 2015MNRAS.447L..21Z} and the extensive BPASS grids \citep[][]{2009MNRAS.400.1019E, 2012MNRAS.419..479E, 2016MNRAS.456..485S} that have been made available to the community. These studies have shown that binary interaction can significantly impact the derived ages and masses of star clusters \citep[e.g.][]{2016MNRAS.457.4296W}, may help to explain the spectral features observed in Wolf-Rayet galaxies \citep{2008A&A...485..657B}, affect star formation rate indicators \citep[e.g.][]{2012MNRAS.424..874L, 2012MNRAS.422..794E} and can significantly boost the ionizing flux \citep[see for example][]{2016MNRAS.456..485S,2016MNRAS.459.3614M,2007MNRAS.380.1098H}. There has also been interest \lang{in} X-ray binaries in \lang{the} context of ionizing radiation \citep{2013ApJ...764...41F}.

One of the challenges for simulations that account for massive binaries is the scarcity of adequate atmospheric models for binary products. Spectral synthesis codes rely on precomputed grids of (1) stellar evolutionary models, which provide information about \lang{key properties} such as evolution of the surface temperature, gravity\lang{,} and composition as a function of time and (2) corresponding atmosphere models, which provide information of the emerging stellar spectrum. For single stars, extensive grids of atmospheric models have been produced to cover the various evolutionary phases \citep{1992IAUS..149..225K,2002A&A...387..244G,1997A&AS..125..229L}. In contrast, the coverage of atmosphere models for stellar objects that are exclusively produced through binary interaction is very sparse or even absent. 

The studies that account for interacting binaries so far have adopted a variety of efficient approximations to \lang{treat atmospheres} of binary products.  These include the usage of atmosphere models originally intended for the evolutionary phases of single stars, possibly after rescaling them.  Other approaches are extrapolation beyond the existing boundaries of the available spectral model grids or the most straightforward approach{color{red}:} simple estimates based on blackbody approximations.

In this work we focus on stars that lose most of their hydrogen-rich envelope through interaction with a companion. This produces very hot, compact helium stars \citep[e.g.][]{1969A&A.....3...83K,1992ApJ...391..246P}. \lang{These stars} are exclusively produced in binary systems. They can be considered as low-mass counterparts to hydrogen-deficient WR stars, i.e.\ helium-burning stars with current masses $\gtrsim 8\Msun$, showing strong broad emission lines indicative of their  strong stellar winds. The WR stars result from the evolution of very massive single stars that lose their envelope through strong winds or eruptive mass loss episodes. In contrast, the progenitors of the less massive stripped stars we are interested in do not have winds that are strong enough to remove the envelope. They can only lose their envelope as a result of interaction with a binary companion. They are related to the \lang{low-mass} subdwarf O and B (sdO/B) stars, which have typical assumed current masses of about 0.5-1\Msun \citep[for a review see][]{2016PASP..128h2001H}. The stripped stars of focus in this study, close the sequence in mass between massive hydrogen-deficient WR stars and the \lang{low-mass} subdwarfs.

Stripped stars represent a common and long-lived evolutionary phase for interacting binaries. Practically every interacting binary produces a hot stripped \lang{star directly} after the first mass transfer phase ceases (if the two stars avoid immediate coalescence). \lang{These stars} are normally powered by central helium burning, \lang{which is} an evolutionary phase that accounts for roughly 10\% of the lifetimes of stars. Their high temperatures and long \lang{lifetimes} along with the expectation that they are common, \lang{make} the stripped stars of interest as potential stellar source of ionizing radiation. Assessing their potential as ionizing sources requires reliable models of their atmospheres. At present, no suitable grids of atmosphere models are available that cover the high effective temperatures and high effective gravities that are typical for these stars.

This paper is intended as first in a series that will systematically explore the structural and spectral properties of stripped stars to evaluate their impact on the spectra of stellar populations. In this first paper we present a case study of the effect of metallicity on a very typical, initially 12\Msun, star that loses its envelope through interaction with a companion after completing its main sequence evolution and before completing central helium burning. This type of mass transfer, case B mass transfer, is the most common case of binary interaction. The specific choice of initial mass is empirically motivated by the observed stripped star in the binary system HD~45166. This allows us to build upon the work by \citet{2008A&A...485..245G} who extensively analyzed and modeled this system. We focus on the long-lived helium burning phase, which is most relevant for the integrated spectrum of stellar populations. 

An additional objective of this study is to improve our understanding of the expected observable characteristics of stripped stars, \lang{which} can be used to observationally identify \lang{these stars}.  Stripped stars are expected to be common, but very few have been observationally identified. The paucity of directly observed stripped stars -- the apparent paradox of missing stripped stars -- can be understood as the result of various selection effects that hinder their detection \citep[][]{2014ApJ...782....7D, abels_paper}. For example, stripped stars are expected to reside near their main sequence companion, \lang{which typically outshines} them in the optical and near UV. Furthermore, \lang{these stars} typically have masses that are too low and orbits that are too wide to cause observable Doppler variation that can be identified easily in the spectra of their companion. \lang{We discuss} spectral features that can be used to overcome these biases. \lang{These spectral features} can be used to guide targeted observing campaigns, \lang{which} will increase the number of observed counterparts in the local Universe. \lang{Observed stripped stars} will provide crucial test cases for the spectral model predictions and simultaneously provide constraints on the outcome of the first phase of binary mass transfer.  

This paper is organized as follows. We describe the evolutionary and spectral models in \secref{sec:modelling}. In \secref{sec:M12} we use our reference model to illustrate the physical processes that determine the structural and spectral properties of stripped stars. We continue by discussing the impact of metallicity on the evolution, the stripping process and the structure of the stripped star in \secref{sec:stripped_Z}. In  \secref{sec:CMFGEN_Z} we discuss the effect of metallicity on the emerging spectra and the spectral features. In \secref{sec:discussion} we discuss the broader implications. We finish with a summary and conclusion in \secref{sec:conclusions}.


\section{Modeling}\label{sec:modelling}

\begin{table*}
\centering
\caption{Physical parameters of our models of stripped stars at different metallicities ($M_{1,\text{init}} = 12\,M_{\odot}$).}
\label{tab:hestar_prop}
{\tiny
\begin{tabular}{lccccccccccccccc}
\toprule\midrule
Model & $Z$ & $M_{\star}$ & $M_{\mathrm{He\,core}}$ & $\log _{10} L$ & $T_{\star}$ & $T_{\mathrm{eff}}$ & $\log_{10} g_{\mathrm{eff}}$ & $R_{\star}$ & $R_{\mathrm{eff}}$ & $\log _{10} \dot{M}$ & $M_{\mathrm{tot, H}}$ & $M_{\mathrm{tot, H, f}}$ & $\tau _{\mathrm{stripped}}$ & $X_{\mathrm{H,s}}$ & $X_{\mathrm{He,s}}$\\ 
 &  & [$M_{\odot}$] & [$M_{\odot}$] & [$L_{\odot}$] & [kK] & [kK] & [cm\,s$^{-2}$] & [$R_{\odot}$] & [$R_{\odot}$] & [$M_{\odot}$\,yr$^{-1}$] & [$M_{\odot}$] & [$M_{\odot}$] & [Myr] &  & \\ 
\midrule 
M\_Z02 & 0.02 & 3.61 & 3.51 & 4.21 & 80.2 & 79.9 & 5.35 & 0.66 & 0.66 & -6.58 & 0.01 & 0.0 & 1.51 & 0.17 & 0.81\\ 
M\_Z0166 & 0.0166 & 3.67 & 3.55 & 4.23 & 79.9 & 79.6 & 5.33 & 0.68 & 0.68 & -6.6 & 0.01 & 0.0 & 1.48 & 0.18 & 0.8\\ 
M\_Z0142 & 0.0142 & 3.71 & 3.57 & 4.24 & 79.5 & 79.1 & 5.31 & 0.7 & 0.7 & -6.63 & 0.02 & 0.0 & 1.46 & 0.19 & 0.79\\ 
M\_Z0134 & 0.0134 & 3.73 & 3.58 & 4.25 & 79.3 & 78.9 & 5.3 & 0.7 & 0.71 & -6.64 & 0.02 & 0.0 & 1.45 & 0.2 & 0.79\\ 
M\_Z01 & 0.01 & 3.81 & 3.63 & 4.27 & 78.6 & 78.3 & 5.28 & 0.74 & 0.74 & -6.68 & 0.03 & 0.0 & 1.42 & 0.21 & 0.78\\ 
M\_Z005 & 0.005 & 4.01 & 3.75 & 4.33 & 76.6 & 76.2 & 5.19 & 0.83 & 0.84 & -6.8 & 0.05 & 0.009 & 1.36 & 0.26 & 0.74\\ 
M\_Z0047 & 0.0047 & 4.02 & 3.76 & 4.33 & 76.2 & 75.8 & 5.18 & 0.84 & 0.85 & -6.81 & 0.05 & 0.011 & 1.35 & 0.26 & 0.73\\ 
M\_Z0021 & 0.0021 & 4.23 & 3.86 & 4.4 & 72.3 & 71.9 & 5.05 & 1.0 & 1.02 & -6.95 & 0.08 & 0.015 & 1.31 & 0.31 & 0.69\\ 
M\_Z001 & 0.001 & 4.41 & 3.92 & 4.45 & 67.4 & 67.0 & 4.89 & 1.23 & 1.24 & -7.09 & 0.12 & 0.02 & 1.29 & 0.36 & 0.64\\ 
M\_Z0005 & 0.0005 & 4.6 & 3.97 & 4.5 & 61.2 & 60.7 & 4.69 & 1.57 & 1.6 & -7.21 & 0.18 & 0.024 & 1.27 & 0.4 & 0.6\\ 
M\_Z0002 & 0.0002 & 4.77 & 3.95 & 4.52 & 54.9 & 54.3 & 4.49 & 2.02 & 2.06 & -8.26 & 0.25 & 0.028 & 1.27 & 0.44 & 0.56\\ 
M\_Z0001 & 0.0001 & 4.89 & 3.92 & 4.53 & 50.3 & 49.7 & 4.34 & 2.43 & 2.48 & -8.4 & 0.31 & 0.031 & 1.26 & 0.48 & 0.52\\ 
\bottomrule 
\end{tabular}
}
\tablefoot{
The parameters displayed here are the initial metallicity, $Z$, the mass of the stripped star, $M_{\star}$, the mass of the helium core of the stripped star, $M_{\text{He core}}$, the luminosity, $L$, the stellar temperature, $T_{\star}$  (at an optical depth $\tau = 100$), the effective temperature, $T_{\text{eff}}$  (at $\tau = 2/3$), the effective surface gravity, $\log_{10} g_{\text{eff}}$, the stellar radius, $R_{\star}$ ($\tau = 100$), the effective radius, $R_{\text{eff}}$ ($\tau = 2/3$), the wind mass loss rate, $\dot{M}$, the total hydrogen mass of the stripped star, $M_{\text{tot, H}}$, the total hydrogen mass at the end of the evolution of the stripped star, $M_{\text{tot, H, f}}$, the time during which the star is stripped, $\tau_{\text{stripped}}$ (the remaining life time after detachment from first Roche lobe overflow phase), the surface hydrogen mass fraction, $X_{\text{H, s}}$ and the surface helium mass fraction, $X_{\text{He, s}}$. Unless stated otherwise we use the parameter values halfway through core helium burning which we define as $X_{\text{He,c}} = 0.5$. The helium core is defined as the most exterior mass coordinate where $X_{\text{H}} < 0.01$ and $X_{\text{He}} > 0.1$. We note that by construction $T_{\star} \approx T_{\text{eff, \textsc{mesa}}}$, the surface temperature resulting from our MESA models. We use the values from our standard wind mass loss models and highlight our reference model in light gray.
}
\end{table*}

\subsection{Stellar evolution with MESA}\label{sec:modelling_MESA}

We \lang{modeled} the evolution of interacting binary stars using the 1D stellar evolution code MESA \citep[version 7624,][]{2011ApJS..192....3P,2013ApJS..208....4P,2015ApJS..220...15P}. We \lang{focused} on stars that lose their hydrogen-rich envelope through Roche-lobe overflow after \lang{they complete} their main sequence and swell to become red giants \citep[also called case~B mass transfer, see][]{1967ZA.....65..251K}. This is the most common case of mass transfer. About a third of all massive stars are estimated to undergo this type of evolution \citep{2012Sci...337..444S}.

We \lang{started} the evolution at the onset of core hydrogen burning using the nuclear network \texttt{approx21}, which contains reactions relevant for hydrogen burning through the CNO cycle and non-explosive helium burning \citep[see][]{2011ApJS..192....3P}. We \lang{followed} the evolution through all long-lived phases until central carbon depletion (defined as $X_{\text{C, c}} < 0.02$). These evolutionary stages together account for 99.9\% of the stellar lifetime.  

We \lang{accounted} for convective mixing using the mixing length approach \citep{1958ZA.....46..108B} adopting a mixing length parameter $\alpha _{\text{MLT}} = 2$. We \lang{allowed} for overshooting above every convective burning region up to $0.335$ pressure scale heights above the convective region, following the calibration by \citet{2011A&A...530A.115B}. The semiconvection parameter is set to $\alpha _{\text{sem}} = 1$ \citep{1991A&A...252..669L}. We also \lang{accounted} for thermohaline mixing \citep{1980A&A....91..175K} and rotational mixing \citep{2013ApJS..208....4P}, but neither of these play a significant role in the models of the mass losing primary star presented here.

We \lang{used} the \cite{2001A&A...369..574V} wind mass loss algorithm for effective temperatures estimated by MESA, $T_{\text{eff, \textsc{mesa}}} > 10^4$~K and surface hydrogen mass fraction, $X_{\text{H, s}}> 0.4$. This prescription scales with metallicity as $\dot{M} \propto Z^{0.85}$. In the models presented here, these conditions are met throughout the main sequence evolution \lang{and} the post-main sequence evolution before the onset of mass transfer.  

After stripping, most of our stars have a surface hydrogen mass fraction $X_{\text{H, s}}< 0.4$ and then we \lang{switched} to the empirically determined WR mass loss algorithm of \cite{2000A&A...360..227N}. This prescription scales with luminosity, surface helium abundance\lang{,} and metallicity, \lang{where} the relation with metallicity \lang{is} $\dot{M} \propto Z^{0.5}$ \citep[cf.][]{1987A&A...173..293K}\footnote{\cite{2015A&A...581A..21H} \lang{have found} a stronger metallicity dependence from observed WR stars, which would imply weaker winds in lower metallicity environments.}. Lacking an appropriate wind mass loss algorithm for the winds of stripped stars, we \lang{extrapolated} this algorithm and \lang{applied} it to the lower luminosities of stripped stars compared to WR stars. The observationally derived wind mass loss rate of the observed stripped star in HD~45166 \citep{2008A&A...485..245G} is in good agreement with the rate estimated by extrapolating the \cite{2000A&A...360..227N} WR algorithm. \lang{This algorithm} almost two orders of magnitude higher than values one would obtain by extrapolating the recent results by \citet{2016A&A...593A.101K} for hot, subluminous stars. We \lang{found} that their mass loss algorithm would give a mass loss rate that is too low to be consistent with the observed spectral characteristics for HD~45166. We therefore chose to adopt \cite{2000A&A...360..227N}. Throughout this work we \lang{considered} variations in the adopted wind mass loss rate for stripped stars to understand the effects of wind mass loss rate.

We \lang{initiated} our models at the zero-age main sequence without rotation. The donor star never reaches high rotation rates throughout its evolution \citep[cf.][]{2010ApJ...725..940Y}. Although we do not discuss the evolution of the accreting star here, we still follow its evolution. During mass transfer the accreting star spins up. We \lang{assumed} the accretion mechanism described in \cite{2005A&A...435..247P} where, when the secondary star rapidly reaches critical rotation during the mass transfer stops accreting significant amount of material. Non-conservative mass transfer follows, \lang{ during which we assumed} that the mass not accreted by the secondary is lost from the system with the specific angular momentum of the orbit of the secondary. The efficiency of mass transfer is uncertain \citep{2007A&A...467.1181D}, but the properties of the primary star are not significantly affected by the details of the treatment of the secondary star and the accretion efficiency.

We \lang{used} the parameters derived for the observed binary system HD~45166 as a motivation for the chosen starting point of our parameter space exploration. HD~45166 consists of a $4.2\,\pm 0.7\,M_{\odot}$ quasi-WR (qWR) star that is orbiting a companion of spectral type B7V. The spectral type of the companion is consistent with a main sequence star of about $4.8 \,M_{\odot}$ \citep{2005A&A...444..895S}. The qWR star is believed to be the stripped remnant of an initially more massive star. We \lang{found} that the parameters of this system can be approximately reproduced adopting $M_{1,\text{init}} = 12\, M_{\odot}$ and  $M_{2,\text{init}} = 5\, M_{\odot}$. The initial mass of the secondary is not as well constrained as the primary. It depends on the efficiency of mass transfer. However, here we are primarily interested in the effect on the primary star. The precise choice of the companion mass does not have large effects on the outcome of the primary star after stripping.  

In this work we \lang{investigated} the effect of metallicity. In a subsequent paper we will discuss the dependence on the further system parameters. We \lang{computed} a grid of binary systems with these initial stellar masses and \lang{adopted} an initial orbital period of 20~days. We \lang{varied} the metallicity between $Z = 10^{-4}$ up to $Z = 0.02$. This corresponds to values of the extremely low metallicity dwarf galaxy IZw18 \citep{1998ApJ...497..227I,2015A&A...581A..15S,2015ApJ...801L..28K} up to super-solar values. The relevant metallicity for HD~45166  is $Z = 0.0166$ \citep[][]{2008A&A...485..245G,2009ARA&A..47..481A}. In terms of [Fe/H] these models span between $-2.18$ and $0.16$, while in terms of A(O) the models span between 6.6 up to 8.9.

For the initial helium mass fraction we \lang{assumed}  $Y = 0.24 + 2 Z$ following \citet{1998MNRAS.298..525P}, which approximately spans the range between an approximately primordial chemistry and a near solar abundance. For hydrogen we \lang{assumed} $X = 1-Z -Y$. The abundances of the heavier elements \lang{were} assumed to scale to solar and meteoric abundance ratios as determined by \cite{1998SSRv...85..161G}. It is likely that the relative metal fraction is not solar for all metallicities. However, for a fixed iron mass fraction we \lang{expected} the effects of CNO variations to be of small impact on the ionizing output, but could affect the strength of carbon, oxygen and nitrogen lines. The initial helium mass fraction could impact the compactness and luminosity of the stars and could be of higher relevance. \tabref{tab:hestar_prop} provides an overview of the main parameters of the evolutionary models presented in this work. 

\begin{figure}
\centering
\includegraphics[width=\hsize]{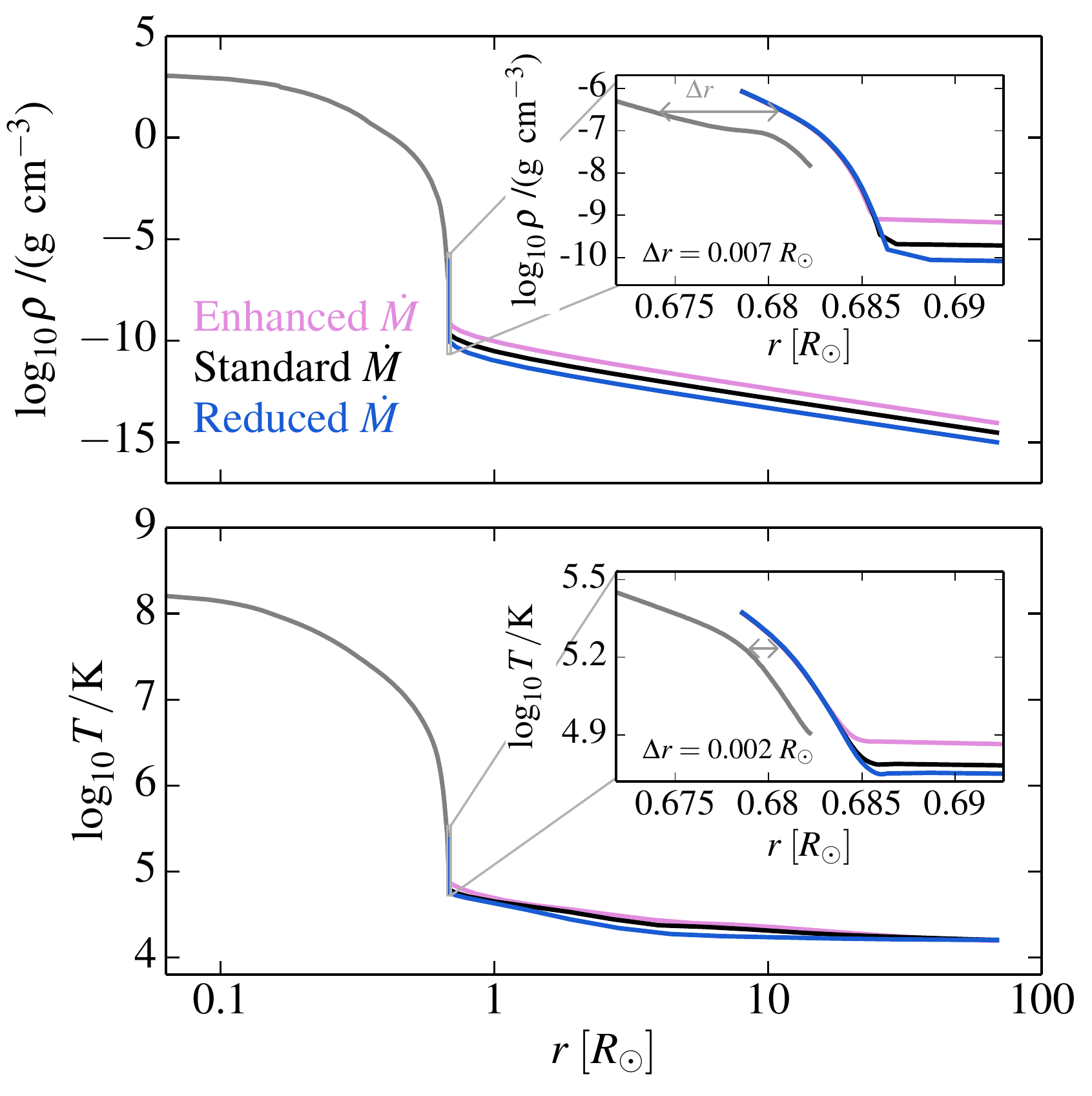}
\caption{Density (top) and temperature (bottom) structure of the MESA (\lang{gray}) and CMFGEN (colored) models of a stripped star and its atmosphere. We show our reference model ($Z = 0.0166$). The different colors of the CMFGEN models correspond to the standard, reduced and enhanced wind mass loss rates. The inset provides a zoom-in, which shows that the connection is not perfect.  However, as shown by the horizontal arrow, the mismatch in the radial direction is very small and translates to an error in the surface temperature that is much less than 1\% (see main text), which is more than sufficient for the purposes of our study.}
\label{fig:stitch_M12}
\end{figure}

\subsection{Stellar atmospheres with CMFGEN} \label{sec:modelling_CMFGEN}

We \lang{used} the radiative transfer code CMFGEN \citep{1998ApJ...496..407H} to model the spectra of stripped stars tailoring them to the MESA stellar evolution models. \lang{The CMFGEN code} takes into account gas out of local thermodynamic equilibrium (non-LTE). \lang{The CMFGEN code} is a necessary tool for modeling optically thick stellar atmospheres, which is something not taken into account by \lang{for example} Kurucz models \citep{1992IAUS..149..225K}. The presence of an extended atmosphere and an optically thick wind changes the optical depth scale, affecting the effective temperature and effective surface gravity (both are defined at a radius where the Rosseland optical depth is $\tau = 2/3$). Each CMFGEN model provides the spectral energy distribution and a normalized spectrum which we \lang{computed} between 50 and 50~000~\AA . We \lang{included} the elements H, He, C, N, O, Si and Fe. Additional elements are not expected to show dominating spectral features as their abundances are low. Our choice of the atomic model is a compromise between computational speed and accuracy.

The observed stripped star in HD~45166 is measured to have the wind speed of $v_{\infty} = 350\,\pm 40\,$km s$^{-1}$ when using a wind velocity beta law with $\beta = 1$ \citep{2008A&A...485..245G}. This wind speed is surprisingly low for a star with $T_{\text{eff}} \simeq 50\,000$~K. There is evidence of a latitudinal-dependent wind in HD~45166 \citep{2008A&A...485..245G}. The spectral lines are consistent with the presence of a fast polar wind ($v_{\infty} \simeq 1\,300\,$km s$^{-1}$) and a slow equatorial wind with ($v_{\infty} \simeq 300\,$km s$^{-1}$), which could reconcile observations and theoretical expectations of hot-star winds. The derived clumping volume filling factor is 0.5 \citep{2008A&A...485..245G}. We \lang{used} these measurements of wind parameters in the qWR in HD~45166 for our spectral models of stripped stars.

Adopting higher terminal wind speeds would decrease the derived optical depth of the wind and also yield broader and weaker lines if the mass loss rate is kept constant. For a constant mass loss rate, lower clumping volume filling factors would increase the equivalent width of recombination lines such as HeII~$\lambda$4686. We discuss the effects of the wind terminal velocity and clumping factors on the spectrum of stripped stars in \appref{app:vinf}.

\subsection{Connecting atmospheres to structure models}\label{sec:tailor}

We \lang{employed} the method described in \cite{2014A&A...564A..30G} to connect the MESA stellar structure with CMFGEN. Here we briefly describe the approach and refer to the paper above for further details. For alternative approaches\lang{,} see \cite{1995ASPC...78..467S}, \cite{2015ApJ...800...97T}\lang{,} and \cite{2017A&A...598A..56M}.

We \lang{took} a structure model from the evolutionary sequence computed with MESA, when the stripped star is halfway through core helium burning, which we \lang{defined} as the moment when the central helium mass fraction drops to $X_{\text{He, c}} = 0.5$. As input for the CMFGEN models at Rosseland optical depth $\tau = 20$, we \lang{used} the effective temperature ($T_{\text{eff, \textsc{mesa}}}$), surface gravity ($\log _{10} g_{\text{\textsc{mesa}}}$), stellar radius ($R_{\text{\textsc{mesa}}}$) and surface abundances extracted from our MESA model following the approach extensively tested by \cite{2014A&A...564A..30G}. From the CMFGEN models we \lang{extracted} the true effective temperature, $T_{\text{eff}}$ ($\tau = 2/3$), and the surface temperature of the star, $T_{\star}$, defined as the effective temperature computed at the radius where the optical depth $\tau=100$. Our standard wind mass loss models have the wind mass loss rate from the \cite{2000A&A...360..227N} WR algorithm, with the exception of the $Z \leqslant 0.0002$ models which are slightly cooler and have the wind mass loss rate of \cite{2001A&A...369..574V}. Given the uncertainties in the mass loss rate, we also computed CMFGEN models with three times higher (labeled ``enhanced'') and three times lower (``reduced'') mass loss rates than our standard values.

\figref{fig:stitch_M12} \lang{shows} the connection between the MESA model and corresponding three CMFGEN models for our reference model ($Z = 0.0166$). The connection is not perfect, but more than sufficient for a reliable prediction of the emerging spectra. The insets in \figref{fig:stitch_M12} highlight the difference $\Delta r$ in stellar radius calculated in MESA and CMFGEN. This discontinuity translates into an uncertainty in the effective temperature of only $\sim100$~K. This is less than 1\% of the surface temperature\lang{,} which \lang{ranges} between 50~000 and 80~000~K (see \tabref{tab:hestar_prop}). Differences of this order are so small that their influence on the predicted emerging spectra an ionizing flux is negligible.


\section{Evolution and spectral features of stripped stars}\label{sec:M12}

The consequences of mass loss due to interaction with a companion in a binary system have been the topic of several classic papers \citep[e.g.][]{1960ApJ...132..146M, 1962AcA....12...28S, 1966AcA....16..231P, 1969A&A.....3...83K, 1973SvA....16..864Y, 1987A&A...178..170V}. We discuss the results obtained with modern higher resolution simulations using updated input physics, \lang{but several} of the physical arguments concerning the evolution can already be found in these early papers. More recent works on this topic includes \cite{2010ApJ...725..940Y}, \cite{2011A&A...528A.131C}, \cite{2013MNRAS.436..774E}, \cite{2015MNRAS.451.2123T}, \cite{2016MNRAS.459.1505M}\lang{,} and \cite{2017ApJ...840...10Y}.

We \lang{used} a 12 \Msun model as an example of the primary in a typical massive binary system. The complete set of adopted parameters are listed and motivated in \secref{sec:modelling_MESA}. We \lang{assumed} an initial separation such that the system evolves through case B mass transfer. This concerns systems \lang{in which} the primary star fills its Roche lobe shortly after leaving the main sequence, typically before the onset of helium burning. Case B mass transfer is the most common type of mass transfer. Moreover, it produces long-lived post-interaction products, which have not yet completed their helium burning phase. This makes the reference model discussed here useful as a starting point for our exploration. \lang{We adopted} a metallicity that is comparable to that found in nearby young star clusters and OB associations in the Milky Way. More precisely, we adopted the metallicity measured for the HD~45166, which is $Z = 0.0166$ \citep{2008A&A...485..245G}. The effects of metallicity are discussed in the next section.

\subsection{Binary evolution and the formation of a stripped star}\label{sec:M12_MESA}

\begin{figure*}
\centering
\includegraphics[width=0.65\hsize]{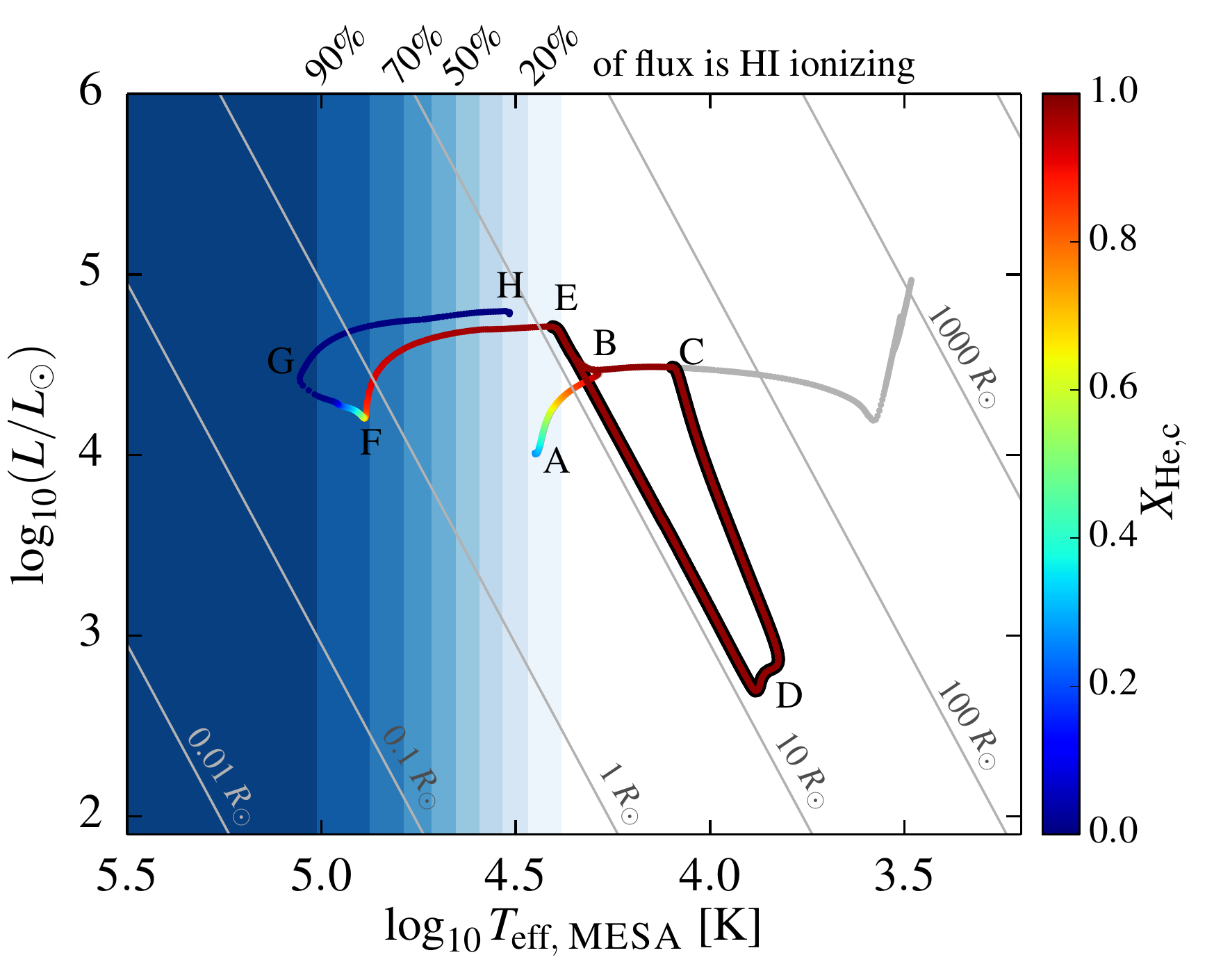}
\caption{Hertzsprung-Russell diagram showing the evolution of a star that loses its hydrogen-rich envelope through Roche-lobe overflow in a binary (our reference model, metallicity $Z = 0.0166$). The color of the track shows the central helium mass fraction, highlighting the long-lasting phases of hydrogen and helium core burning with color change. The thicker, black background line \lang{indicates} the Roche-lobe overflow. Letters \lang{denote} different evolutionary \lang{stages described} in the text. At point F the central helium mass fraction \lang{of the stripped star} decreases and passes $X_{\text{He, c}} = 0.5$ -- the point when we model the spectrum (see \secref{sec:M12_CMFGEN} and \figreftwo{fig:SED_M12}{fig:spectra_M12}). The gray track is that of a corresponding 12\Msun single star. The blue background shading shows the fraction of emitted flux that is hydrogen ionizing, assuming blackbody radiation. \lang{The gray} straight \lang{lines indicate} lines of constant radius.}
\label{fig:HRD_M12}
\end{figure*}

In \figref{fig:HRD_M12} we show the evolutionary track computed with MESA of a 12 \Msun single star in \lang{gray}, together with the evolutionary track for the primary star in our binary reference model. Initially, the two stars evolve similarly as they move along the main sequence (labeled A-B). The central helium abundances $X_{\text{He, c}}$ (plotted in color) steadily \lang{rise} as the both stars fuse hydrogen into helium in their convective cores.  

\begin{figure*}
\centering
\includegraphics[width=0.5\hsize, trim=0 7cm 0 7cm, clip]{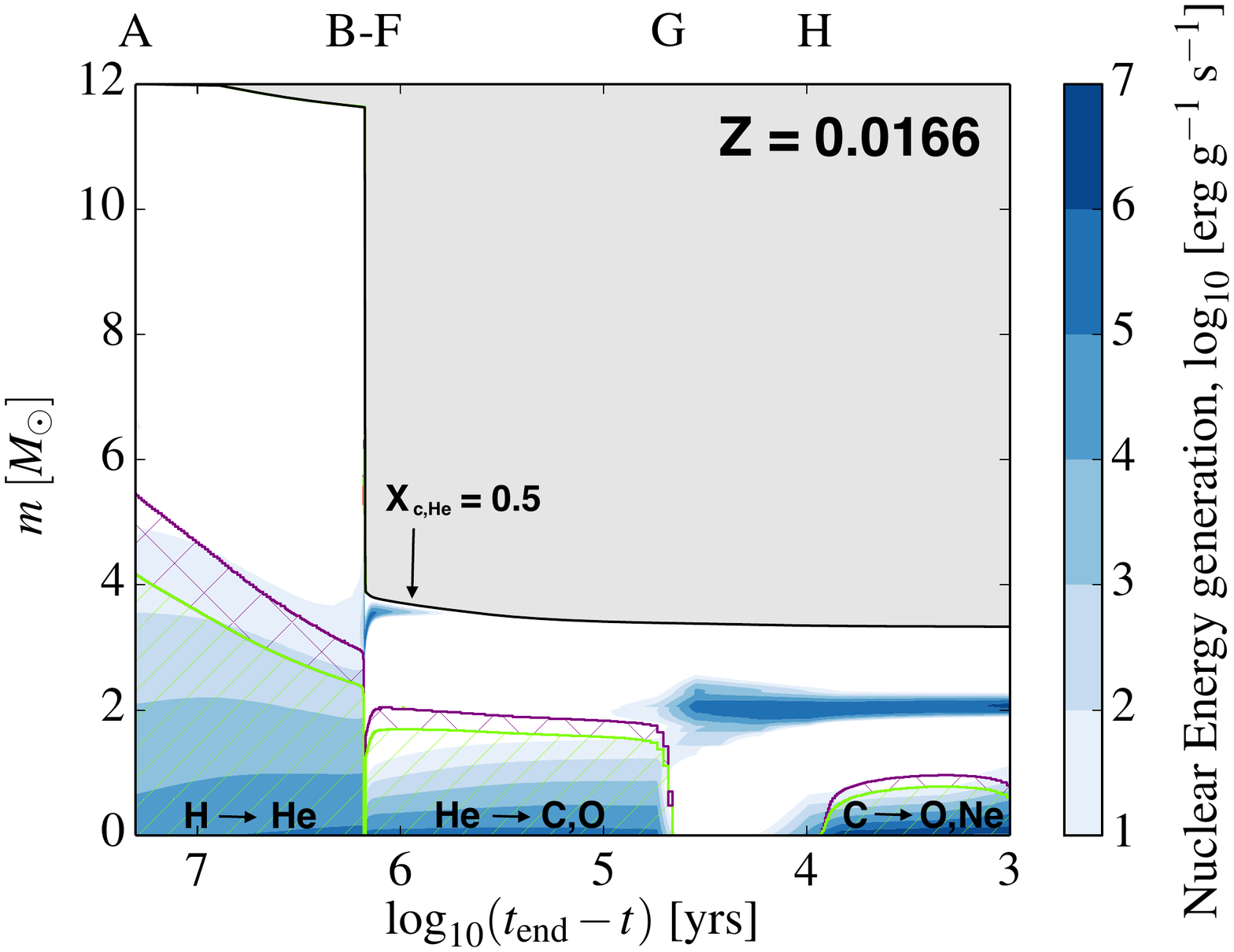}%
\includegraphics[width=0.5\hsize, trim=0 7cm 0 7cm, clip]{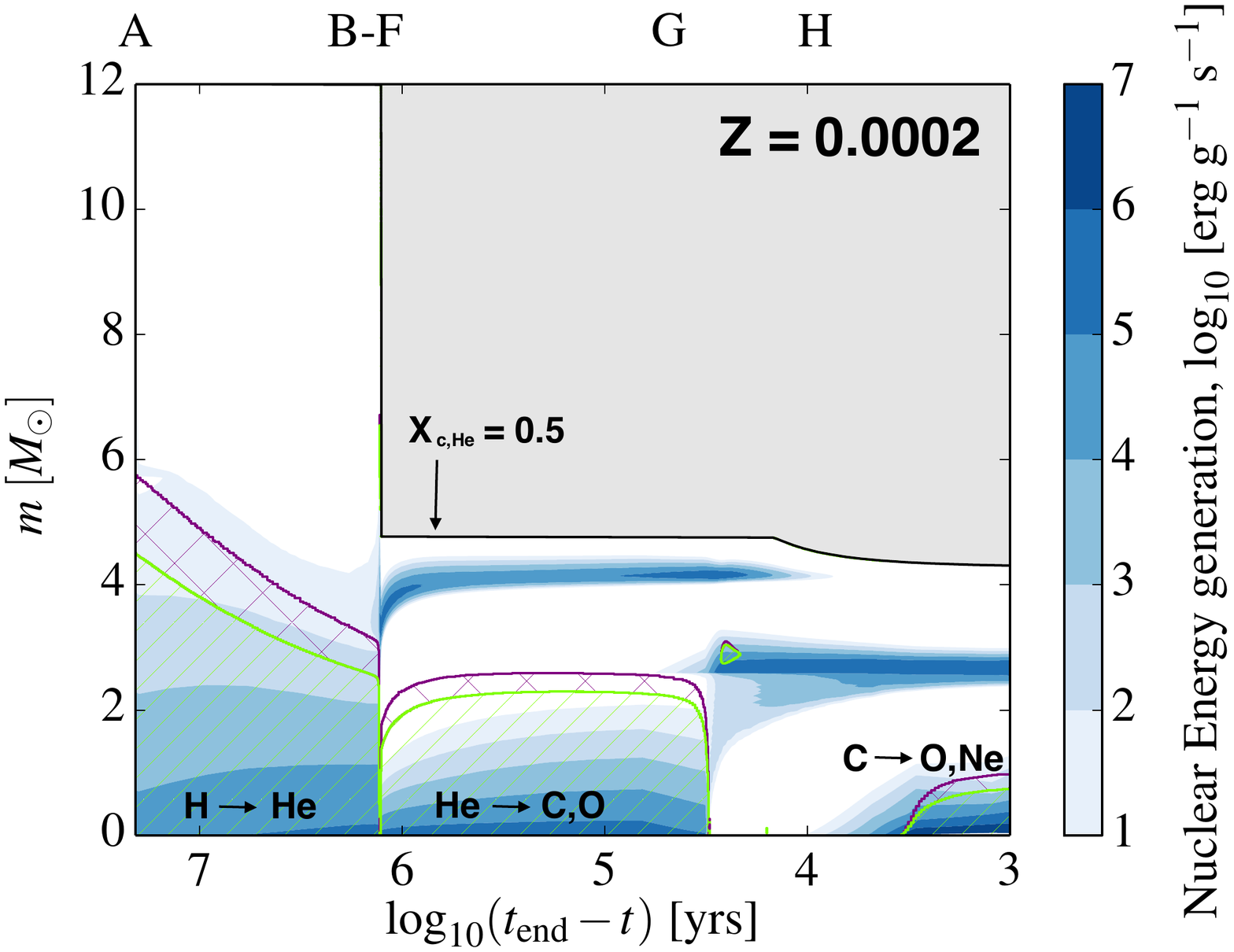}
\caption{\textit{Left panel:} Kippenhahn diagram showing the interior structure of the $12\; M_{\odot}$ primary star of the reference model over time. We show nuclear burning regions shaded in blue, convection in hatched green and overshooting in crosshatched purple. The solid black line shows the surface of the star, which decreases quickly when the envelope stripping occurs. The letters \lang{denote} the same evolutionary phase as labeled in \figref{fig:HRD_M12}. \textit{Right panel:} Corresponding Kippenhahn diagram for the lower metallicity ($Z = 0.0002$) model. \lang{There is a} larger amount of \lang{leftover} hydrogen-rich envelope after Roche-lobe overflow.}
\label{fig:Kipp_M12}
\end{figure*}

After about 18.7~Myr both stars have exhausted their central fuel. Nuclear burning of hydrogen continues in a shell around the helium core and both stars expand. The single star expands freely to become a red supergiant, reaching a final radius in excess of 1000 \Rsun. In contrast, the primary star in our binary model fills its Roche lobe at point C and starts to rapidly lose mass. This can be seen in the left panel of \figref{fig:Kipp_M12}, where we plot how the mass changes as a function of time. We also show the mass coordinates of the regions \lang{in which} nuclear burning takes place (blue shading) and the interior regions that are mixed by convection and overshooting (green diagonal lines and purple cross hatching, respectively).

The primary star has a radiative envelope at the onset of mass transfer. As the outer, highest entropy layers are removed, the envelope initially responds on a dynamical timescale by shrinking. On a longer, thermal timescale\lang{,} the star is still trying to expand and cross the Hertzsprung gap leading to continued stable mass transfer. The Roche lobe limits the size of the star.  

During the mass transfer phase the luminosity drops by more than an order of magnitude. This is because the deeper layers of the star need to expand as they adjust to the quickly decreasing total mass. The energy needed for this causes a brief, large drop in the luminosity. This continues until the maximum mass loss rate is reached (D). Once the mass loss rate starts to decrease again, the readjustments in the thermal structure require less energy so luminosity can again increase.  

At point E the donor star has lost more than 8 \Msun. The surface layers that are now exposed are helium-rich layers that were once part of the convective core. The star detaches and mass transfer stops. The now stripped core of the primary contracts and heats up until thermal equilibrium is restored at point F. The central regions have now reached temperatures \lang{that are} high enough to ignite helium burning, which burns in the convective core to carbon and oxygen. 

\subsection{Characteristics and evolution of the stripped star}
 
Our main phase of interest is the helium burning phase of the stripped star (F). This is the longest lasting evolutionary phase after the main sequence, accounting for almost 10\,\% of the total lifetime in our model. Given the high fraction of young stars in close binaries \lang{that undergo} similar evolution, one would expect on average a few percent of all stars to reside in this phase at any given time. 

Despite having lost over \lang{two-thirds} of its mass, from 12 \Msun initially down to 3.6 \Msun after stripping, the bolometric luminosity, $\log_{10} (L/\Lsun) \sim 4.2$, is still comparable to the luminosity it had during its main sequence evolution. For an overview of the properties, see \tabref{tab:hestar_prop}. The star is very compact  $\sim 0.7\, \Rsun$, has a very high surface gravity, $\log_{10} g_{\rm eff} \sim 5.3$\lang{,} and is very hot.

With an effective temperature $\sim$80~000~K, the radiation emitted by the stripped star is expected to peak in far/extreme UV. In \figref{fig:HRD_M12} we indicate the fraction of photons emitted at wavelengths blue-ward of $\lambda \leqslant 912$ \AA\ (Lyman continuum photons) by a blackbody with the corresponding effective temperature, with blue vertical contours. A blackbody with a temperature of 80~000~K, similar to the effective temperature of our stripped star, emits about 85\,\% of its radiation in Lyman continuum photons. For comparison, the equivalent single star model (shown in \lang{gray} in \figref{fig:HRD_M12}) has an effective temperature of only  $\sim$3~000~K during its helium burning phase. A single star of this mass does not emit \lang{a} significant number of ionizing photons throughout its entire life. 

Initially, we find that a thin shell containing a mixture of hydrogen and helium remains at the surface. \lang{This shell} contains less than 0.1\Msun of hydrogen. This is sufficient to sustain a weak hydrogen burning. The burning shell quickly moves outward in mass coordinate as it converts H into He as can be seen in \figref{fig:Kipp_M12}. At the same time mass is lost from the surface by the stellar wind at a rate of $\dot{M} \sim 2.5 \times10^{-7}\Msun\,{\rm yr}^{-1}$ in this model. This is high enough to remove the hydrogen layer by the time core helium depletion is reached. See also panel e of \figref{fig:evol_strip}, which shows the evolution of the surface abundance of hydrogen and helium as a function of time. 

When central helium burning ceases, the entire star contracts and heats up to point G, where the helium shell ignites. The C/O core continues to contract with the helium burning shell on top. The envelope responds by expanding until the star reaches balance at point H. During the expansion, carbon and later also oxygen ignite in the core. We \lang{stopped} the computations at central carbon depletion. The following phases of the evolution are so fast that the luminosity and effective temperature \lang{do} not have time to change significantly any more. With a helium layer of about $1.3\,M_{\odot}$ and almost pure helium on the surface, the star is expected to end its life as a H-deficient (type Ib) supernova.

 \begin{figure*}[t]
\centering
\includegraphics[width=0.9\hsize]{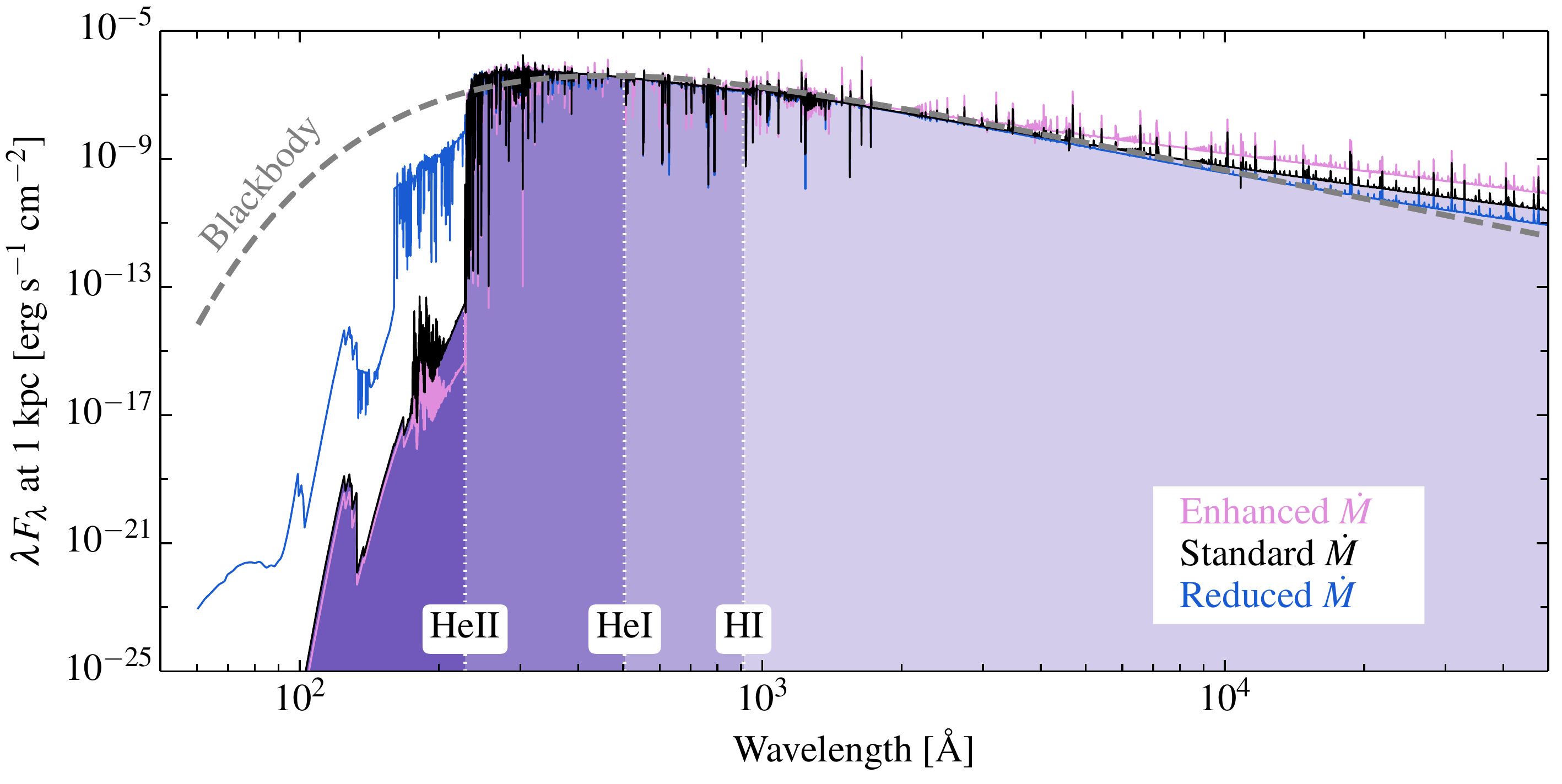}
\caption{Spectral energy distribution of our reference model with standard mass loss rate in black together with the blackbody with temperature same as $T_{\text{eff,\textsc{mesa}}}$ \lang{(dashed gray line)}. In pink and blue we show the reduced and enhanced mass loss rate models. We highlight the high rate of ionizing flux emitted from this star in purple shading.}
\label{fig:SED_M12}
\end{figure*}

\begin{figure}
\centering
\includegraphics[width=\hsize]{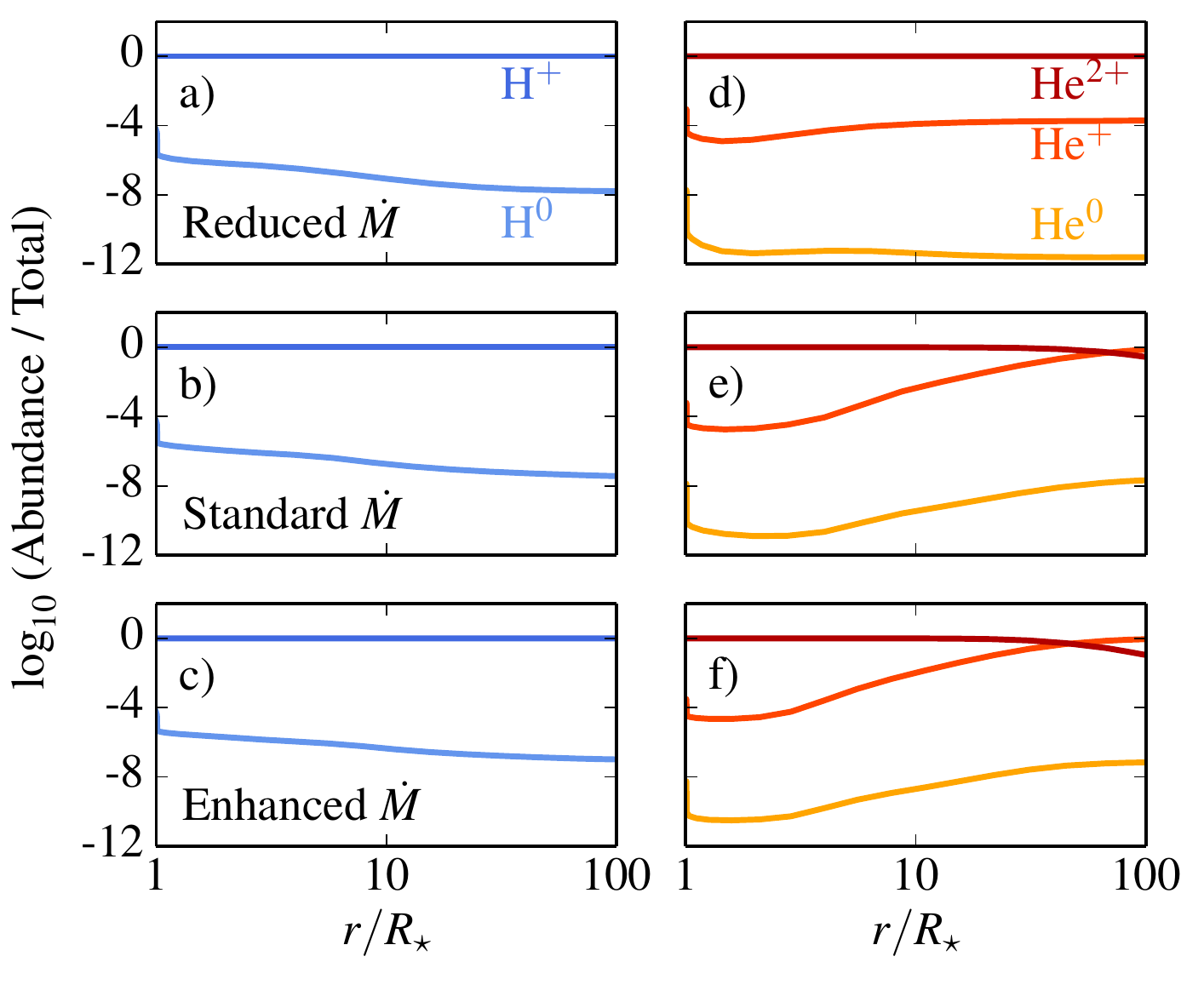}
\caption{\lang{Ionization} structure of hydrogen (left) and helium (right) in the three atmosphere models of our reference model. From top to bottom we show the reduced, standard and enhanced mass loss rate models.}
\label{fig:ionisation_structure_M12}
\end{figure}

\subsection{\lang{Resulting} spectrum of a stripped star}\label{sec:M12_CMFGEN}

\begin{figure*}
\centering
\includegraphics[width=\textwidth]{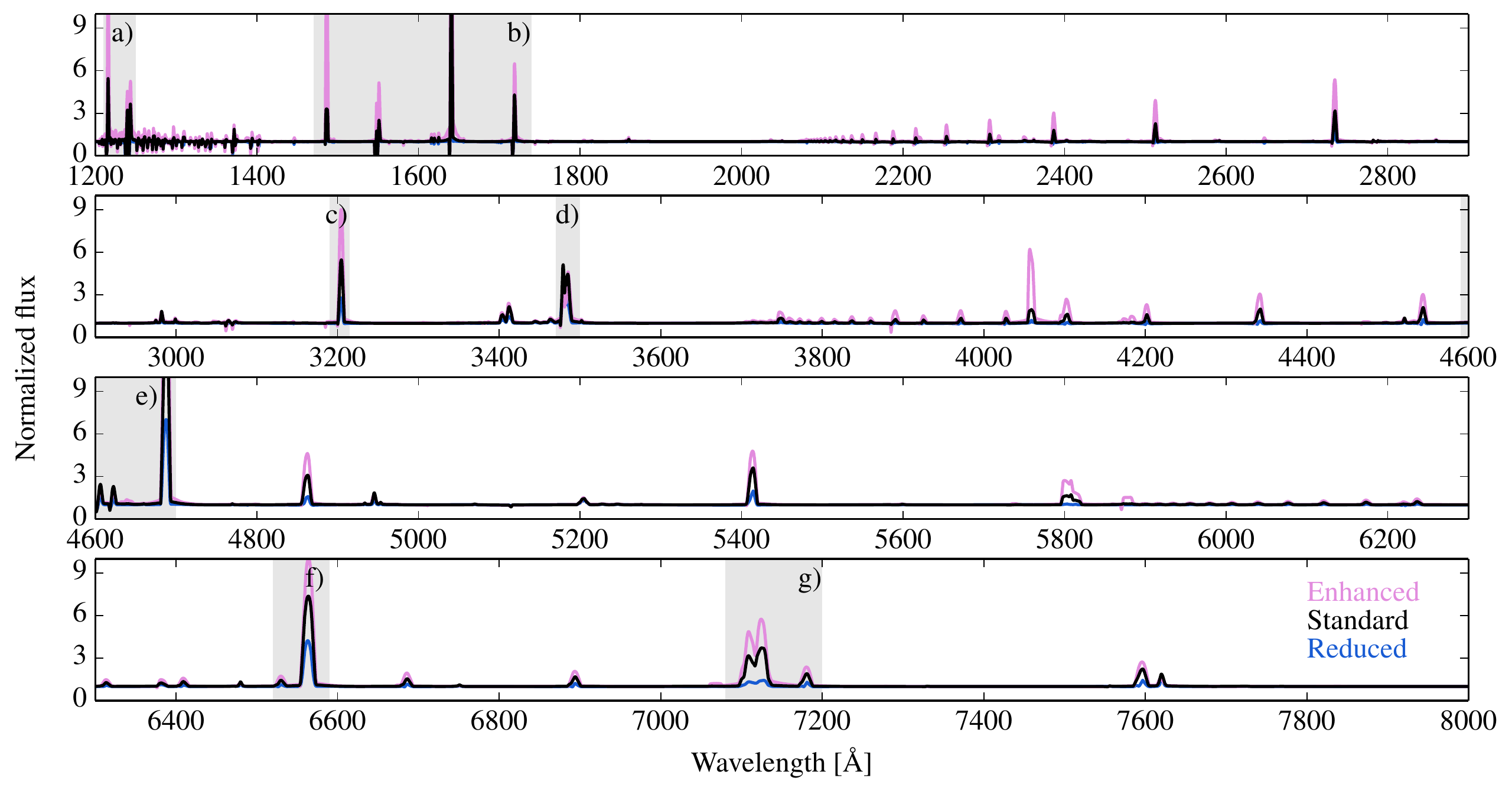}
\includegraphics[width=\textwidth]{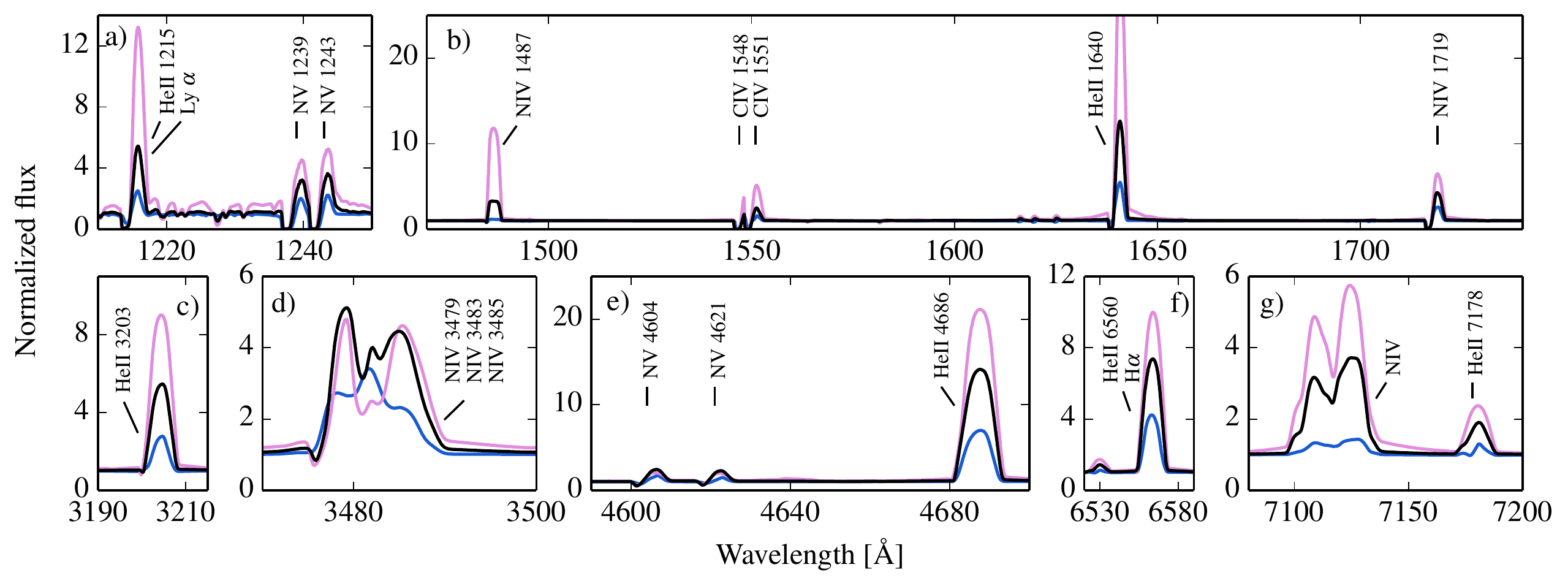}
\caption{CMFGEN spectra for our reference model. We show the standard, enhanced and reduced mass loss rate models in black, pink and blue respectively. The \lang{gray} shaded regions are shown as zoom-ins in the bottom panels.}
\label{fig:spectra_M12}
\end{figure*}

We \lang{computed} a tailor-made atmosphere model with CMFGEN for the stripped star computed with MESA at the moment when it is halfway through helium burning, using the procedure described in \secref{sec:tailor}. \figref{fig:SED_M12} shows the resulting spectrum (black line). The wavelengths corresponding to photon energies required to ionize neutral hydrogen (HI), singly ionize helium (HeI)\lang{,} and fully ionize helium (HeII) are \lang{indicated} (vertical dotted white lines). For comparison, we overplot a blackbody spectrum for a temperature of $\sim$80~000~K (\lang{gray} dashed line), \lang{corresponding} to $T_{\text{ eff, \textsc{mesa}}}$, which is the computed effective temperature resulting from the MESA model. 

The blackbody approximation strongly overestimates the extreme UV flux (\lang{shortward} of 228$\,\AA$) and underestimates the flux at peak\lang{, which occurs between 228-512$\,\AA$}. It also underestimates the emission at longer wavelengths\lang{, these are most clearly visible in \figref{fig:SED_M12} for wavelengths longer than about 4000$\,\AA$, where reprocessed ionizing photons are emitted at longer wavelengths in the form of recombination lines.} The excess in infrared and radio wavelengths compared to the blackbody is due to free-free emission. The sharp drop at $\lambda = 228$\,\AA\, is due to \lang{the} recombination of He$^{2+}$ to He$^{+}$ in the outer parts of the wind, allowing for He$^{+}$ to absorb a significant fraction \lang{of high-energy} photons. We do not observe a similar drop at the hydrogen ionizing limit (912\,\AA ), since the hydrogen is completely ionized throughout the wind. 

In \figref{fig:ionisation_structure_M12}, we show the ionization structure of hydrogen and helium as a function of the distance from the stellar surface. For helium we find a helium recombination front at a distance of about 80 stellar radii (panel e). For our standard model, we find that hydrogen is fully ionized out to a distance of about $100\,R_{\star}$, which is the computational domain considered (panel b). At larger distances the density drops further and therefore we do not expect hydrogen to recombine. This assures that no major changes to the spectrum \lang{are} caused by the wind outflow at larger distances than considered in our model. 

Varying the adopted mass loss rate in the CMFGEN models has a large effect on the extreme UV flux as can be seen in \figref{fig:SED_M12}. Increasing the mass loss rate enhances the density in the wind, effectively making it more optically thick in the continuum. This moves the photosphere outward and reduces the effective temperature. In the case of our model with enhanced mass loss, we find that the photosphere is located  at 0.99 \Rsun, while the stellar surface is located at 0.68 \Rsun.  The resulting effective temperature, $T_{\rm eff} \sim$70~000~K, is almost 10~000~K lower than the temperature at the stellar surface. This effect is commonly observed for the the higher mass counterparts to stripped stars, WR stars, which typically have dense, optically thick winds. For the models that adopt our standard and reduced mass loss rates we find no significant difference between the effective and the surface temperature, indicating that in these cases the wind is optically thin in the continuum. Many spectral lines are however optically thick, as we discuss in \secref{sec:spectral_features_M12}.

Considering the relation between wind mass loss rate and radiation pressure through the Eddington factor ($\Gamma_e$), we expect a realistic wind mass loss rate in the considered parameter range to be closer to \lang{that} used in the "reduced" mass loss model \citep{2014A&A...570A..38B}. However, these estimates \lang{were derived} from more massive stars and therefore also \lang{included} an extrapolation to the lower mass regime of the presented stripped stars.

\subsection{Characteristic spectral features}\label{sec:spectral_features_M12}

To show the spectral features more clearly, we plot the normalized spectra in \figref{fig:spectra_M12} for our standard, enhanced and reduced mass loss \lang{rates}. Our CMFGEN models show emission lines with parabolic \lang{shapes}, which is typical \lang{for} lines \lang{that are created in the wind and are optically thick}. 

The strongest line in the optical spectra is the HeII~$\lambda$4686 with an equivalent width of 95.6\,\AA\ and with flux at the peak corresponding to more than \lang{10} times that of the neighboring continuum in the standard wind mass loss model (see enlargement in panel e of \figref{fig:spectra_M12}). 

Other strong lines in the optical band are the blend of HeII~$\lambda$6560 and H$\alpha$ (panel f) and a mix of NIV lines in the range 7100-7140~\AA\ (panel g). We also find weaker emission in CIV~$\lambda$5801 and CIV~$\lambda$5812. In the enhanced mass loss model HeI~$\lambda$5875 \lang{shows up owing} to the higher abundance of He$^{+}$ in the wind compared to the lower mass loss models. The \lang{strongest line of the UV spectrum} is HeII~$\lambda$1640, but also the Ly$\alpha$ and HeII~$\lambda$1215 blend, NIV~$\lambda$1487 and NIV~$\lambda$1719 are strong in emission. In the infrared we find the strongest lines to be HeII~$\lambda$18636, $\lambda$30908 and $\lambda$47622. 

Varying the mass loss rate affects the strength of the emission lines. Lower mass loss rates consistently show weaker lines. If stripped stars had \lang{mass loss rates that were} even lower than those presented here, our models indicate that they would show little or negligible emission, and mostly an absorption-line spectrum.

\lang{Stripped} stars share common characteristics with central stars in planetary nebulae (CSPNe) \citep{2011MNRAS.414.2812D,  2011A&A...526A...6W}. However, they typically have very high surface gravity, implying broader lines compared to stripped stars. We also expect them to generally be of lower mass and post central helium burning objects. 


\section{Effect of metallicity (I): \lang{The} formation and structure of stripped stars}\label{sec:stripped_Z}

Numerous studies have discussed the effect of metallicity on the structure and evolution of single stars \citep[e.g.][]{1965ApJ...141.1019B, 1970A&A.....5...12C, 1986A&AS...66..191B, 1990A&AS...84..139M, 1992A&AS...96..269S, 1993ApJS...88..509C, 2000A&AS..141..371G, 2001A&A...373..555M, 2008A&A...490..769C, 2011A&A...530A.115B, 2013A&A...558A..46P, 2015A&A...581A..15S}. Summarizing these works, we can distinguish three main effects of metallicity on (1) the opacity, most notably in the subsurface layers, (2) the nuclear burning rate, especially of hydrogen fusion through the CNO cycle\lang{;} and (3) the mass loss rates for radiatively driven winds. Here, we describe how the different effects of metallicity interplay in case of the mass losing star in a binary system. We consider metallicities between $Z = 0.0001$ and $Z=0.02$. \lang{In the sections below, we} discuss the consequences for the pre-interaction phase, the removal of the envelope, the resulting characteristics of the helium star\lang{,} and its emerging spectrum. 

\begin{figure*}
\centering
\includegraphics[width=0.75\hsize]{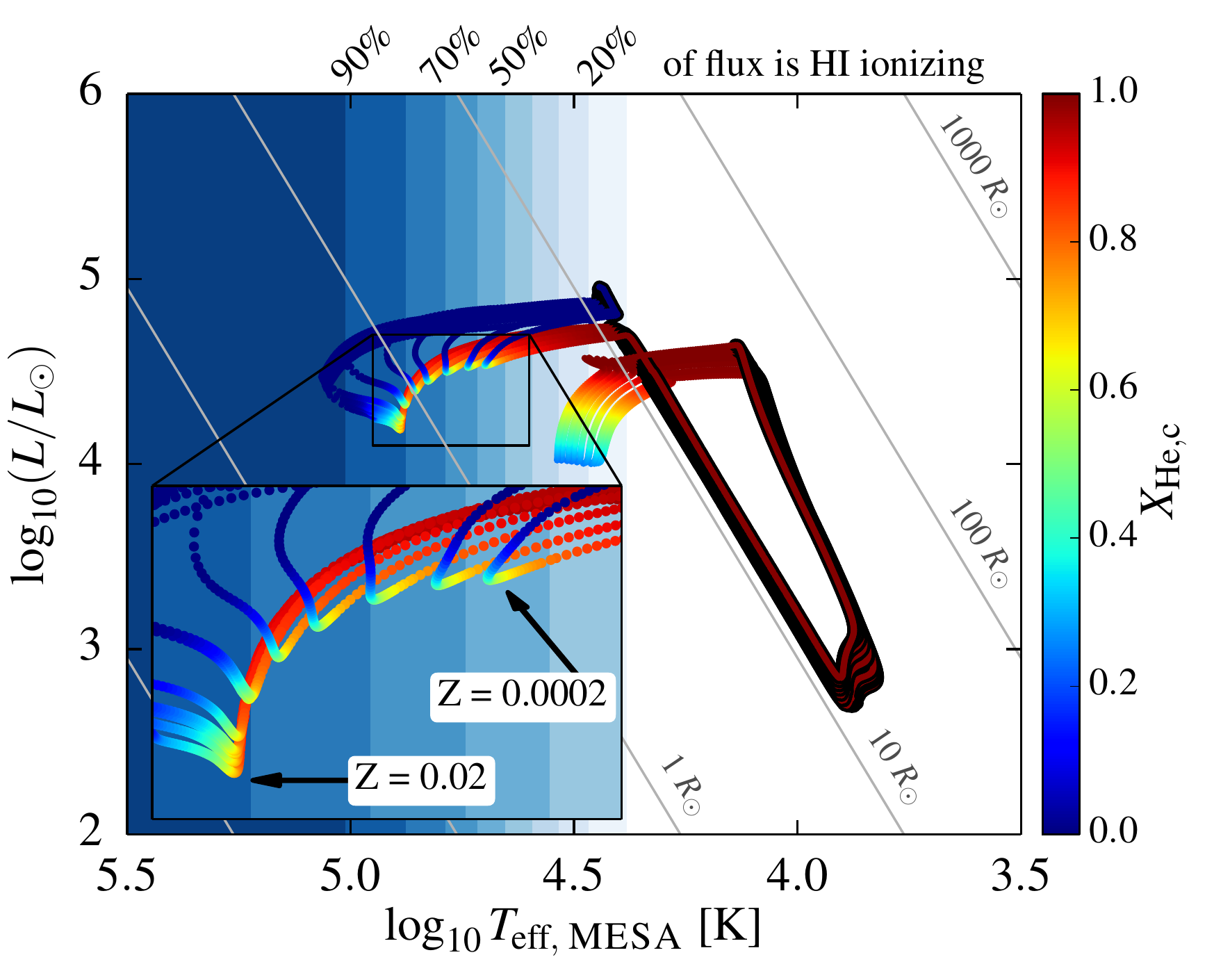} 
\caption{Evolutionary tracks of donor stars with metallicity ranging from $Z = 0.0001$ up to $Z = 0.02$ (similar to \figref{fig:HRD_M12}) as they evolve from the zero-age main sequence until they experience mass loss by Roche-lobe overflow (black outlined). Their subsequent evolution as stripped stars is highlighted in the inset panel. The low metallicity models are more compact and hotter before interaction. However, after mass transfer they are cooler and more luminous than the high metallicity stripped stars. \lang{A} discussion of the evolution \lang{is provided in} \secrefthree{sec:evol_Z_pre}{sec:evol_Z_rlof}{sec:evol_Z_strip}.}
\label{fig:HRD_Z}
\end{figure*}

\subsection{The pre-interaction phase}\label{sec:evol_Z_pre}

During the pre-interaction phase both stars effectively evolve as single stars. Throughout the main-sequence evolution, we find that metal-poor stars are hotter, more compact and a little more luminous, as can be seen in \figref{fig:HRD_Z}. This is consistent with what has been found for single stars as described in the studies mentioned above.  

This is, in part, due to the effect of metallicity on the nuclear reaction rates for hydrogen burning through the CNO cycle, as already pointed out in the earliest papers on the subject \citep[cf.][]{1965ApJ...141.1019B}. At lower metallicity, the catalysts for the CNO cycle are more scarce and hydrogen burning is less efficient. The star compensates by contracting, which increases the central temperature. This, in turn, raises the nuclear reactions rates to \lang{levels that are} required for the star to be in thermal equilibrium. The result is a hotter, more luminous star. 

These days, stellar evolutionary calculations adopt more realistic opacity tables \citep[e.g., OPAL,][]{1996ApJ...464..943I}, which have a more sophisticated treatment of the many bound-bound and bound-free transitions due to metals. These are most important in the cooler subsurface layers of the star, where heavy elements are only partially ionized. The most prominent example is the so-called iron peak, which plays a role in the \lang{subsurface} layers at $\log_{\rm 10} T \sim 5.2$ \citep{2009A&A...499..279C}. The lower opacity in our metal-poor models \lang{contributes} to the fact that they are hotter and more compact. 

In the deep interior, electron scattering is the dominant source of opacity, which does not depend on the metallicity, at least not directly, as $\kappa_{\rm es} = 0.2 \, (1+X_{\rm H}) \, {\rm cm}^2 \,{\rm g}^{-1}$. However, when initializing our models we chose to scale the initial mass fraction of hydrogen and helium with metallicity. The hydrogen abundances decrease as the metallicity rises in agreement with the overall trend expected for the chemical enrichment of galaxies over cosmic time. This way we introduced a mild indirect dependency of $\kappa_{\rm es}$ on the metallicity. Our metal-poor models \lang{initially have} a $\sim 8$~\% higher value for $X_{\rm H}$ than our metal-rich models.

The effect on the opacity in the interiors is small, but of importance since the electron scattering plays a role in determining the extent of the convective core, which in turn determines how much fuel the central burning regions can access. \figref{fig:Kipp_M12} shows the extent of the convective core in mass coordinate. Our metal-poor model \lang{indeed has} a more massive convective core at the start of its evolution \lang{because of} a combination of the effects discussed above.  

We can further see the effect of the metallicity dependence of mass loss by radiatively driven stellar winds. Our metal-rich model loses a few percent of its mass over the course of the main sequence evolution (labeled A-B in \figref{fig:Kipp_M12}). Our low metallicity model does not show any significant mass loss by stellar wind.  

Both the extent of the convective core and stellar wind mass loss have consequences for the final mass of the helium core when the star leaves the main sequence $M_{\text{He core, TAMS}}$. We find a small difference in mass of about 6~\%, with the metal poor star having the more massive core ($M_{\text{He core, TAMS}}$ = 3.14 and 2.95 \Msun, respectively) in addition to subtle differences in the chemical profile above the helium core. The slightly larger core mass and larger total mass for our metal-poor models explain why our metal-poor stars are substantially brighter when they leave the main sequence as can be observed when comparing the location of the characteristic hook feature that marks the end of the main sequence (see \figref{fig:HRD_Z}).  

\begin{figure}
\centering
\includegraphics[width=\hsize]{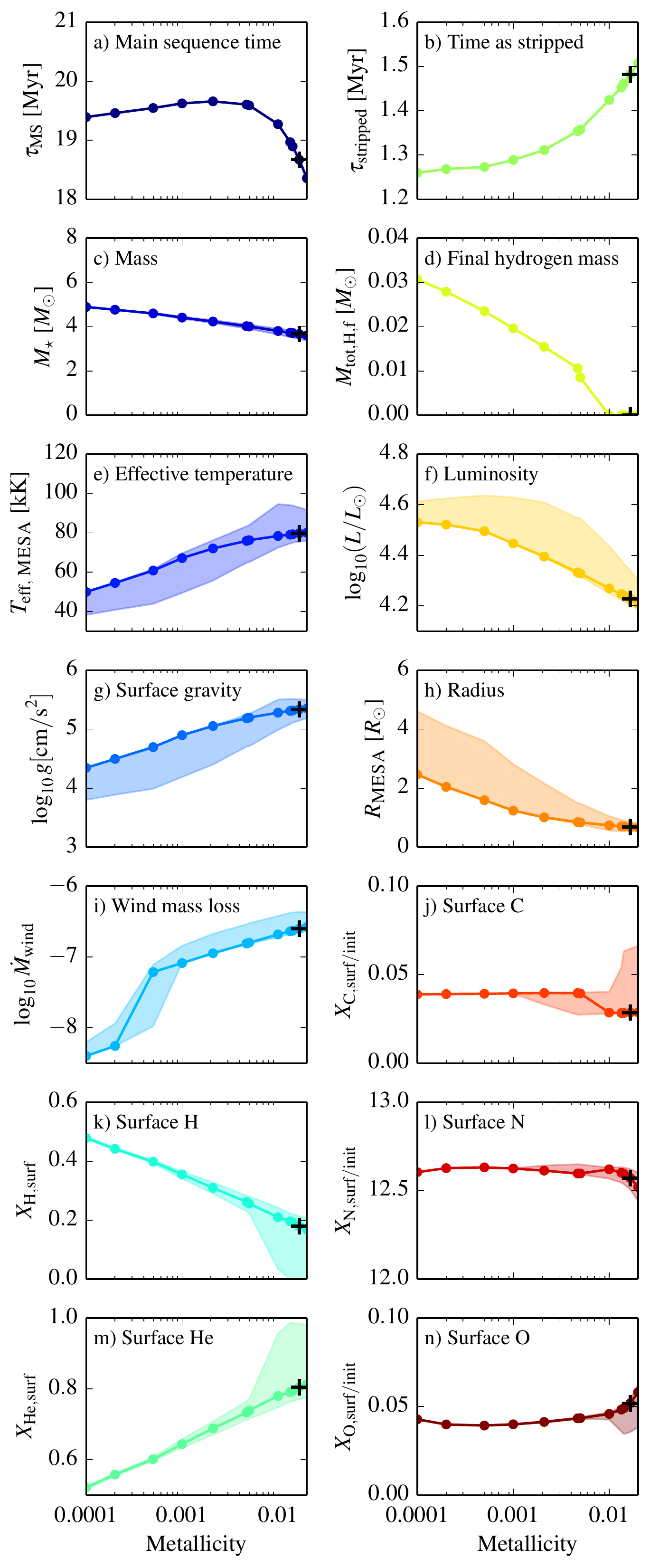} 
\caption{Properties of the evolutionary models of stripped stars (see \tabref{tab:hestar_prop}) shown as function of metallicity. The markers and connecting lines highlight when the central helium abundance is $X_{\text{He, c}}= 0.5$, these values are given in \tabref{tab:hestar_prop}. The shaded areas indicate how each model \lang{evolves} from core helium mass fraction of 0.9 to 0.05 during core helium burning. We highlight the model with metallicity $Z = 0.0166$ with a black plus sign (described in \secref{sec:M12}).}
\label{fig:MESA_hestar_prop}
\end{figure}

In \figref{fig:MESA_hestar_prop} (panel a) we compare the main sequence \lang{lifetimes of the various metallicity models}. Reducing the metallicity from 0.02 down to $\sim$0.0021 increases the lifetime by about 8\%, owing to the amount of nuclear fuel that the star can access. When we reduce the metallicity further to $0.0001$, this effect \lang{is saturated}. The various effects that lead to a higher luminosity start to dominate. The stars effectively burn faster through their available fuel. We find that the main sequence lifetime slightly decreases again.

When the star leaves the main sequence it continues to burn hydrogen in a shell around the core. The shell briefly drives a convection zone, which influences the details of the shape of chemical profile above the core, but we find that it has no influence on the stripping process for the models presented here. For our most metal-rich model, we find that a region of about 1\Msun above the core is partially burned, when the star leaves the main sequence. For our most metal-poor model we find that the region extends to about 1.5\Msun above the core. 

\subsection{\lang{Onset} of Roche-lobe overflow and the removal of the envelope}\label{sec:evol_Z_rlof}

The stars we consider here fill their Roche lobe shortly after leaving the main sequence during the hydrogen shell burning phase. Throughout their pre-interaction evolution, the metal-poor models have been more compact. They need to expand further in order to fill their Roche lobe. Effectively, they fill their Roche lobe at a slightly more advanced evolutionary stage. 

The size of the Roche lobe is approximately the same at the moment the primary fills its Roche lobe for the first time, about 27\Rsun in all our simulations. There is a small difference because of the \lang{metallicity-dependent} stellar wind mass loss rates. Mass loss from our metal-rich systems in the form of fast stellar winds leads to widening of the orbit and thus increasing the Roche lobe. This effect is partially compensated by the reduction of the mass ratio, $q = M_2/M_1$, which reduces the relative size of the Roche lobe of the donor. The net result is that the metal-rich models are slightly larger (2~\%) at the time they fill their Roche lobe. However, we do not expect this to play a role of significance. When conducting test experiments varying the binary separation and mass ratio we find variations in outcome that are negligible.  
  
The process of mass stripping by the companion (marked with a black outline in \figref{fig:HRD_Z}) proceeds \lang{in a similar way} in all models\lang{;} see \secref{sec:M12_MESA}. Initially, \lang{the stars} respond by contracting rapidly on a dynamical timescale and they subsequently expand on the slower thermal timescale of the outer layers. \lang{However, the dynamical phase is not modeled explicitly, but enforced by the assumption of hydrostatic equilibrium.} The mass transfer rate peaks at slightly higher values in the metal-poor models. Eventually, when most of the envelope is removed and helium enriched layers are exposed to the surface, there is a certain point \lang{at which} the star is no longer able to expand in response to any further mass removal. At this point the stars detach as the donor starts to contract. 
 
For our metal-rich models, we find that a larger amount of mass removal is needed to reach this point. Our metal-poor donor stars are still about 4.9\Msun when they detach, more than a solar mass more than our metal-rich models which are only about 3.8\Msun at this moment. For our metal-rich models the helium surface mass fraction at the moment the two stars detach is about 0.75\lang{; this compares} to about 0.55 in our metal-poor models.  

We expect that the reduced opacity in the outer layers of the metal-poor models is an important factor. In the subsurface layers where the iron peak plays a role, we find that the opacity is about three times higher in our metal-rich models. In addition, also the subtle differences in the interior chemical profile that are inherited from the pre-interaction evolution play a role. A further effect that may contribute is the difference in central temperature. The metal-poor models have more massive and more compact cores, and thus higher central temperatures, which allows for the ignition of central helium burning before the star detaches from its Roche lobe. Our most metal-rich model is still primarily powered by H shell burning when it detaches. 

After the stars detach we find that all models contract, fully ignite helium in their center if they had not already done so, and settle to their thermal equilibrium structure as a central helium burning star.

\begin{figure}
\centering
\includegraphics[width=0.8\hsize]{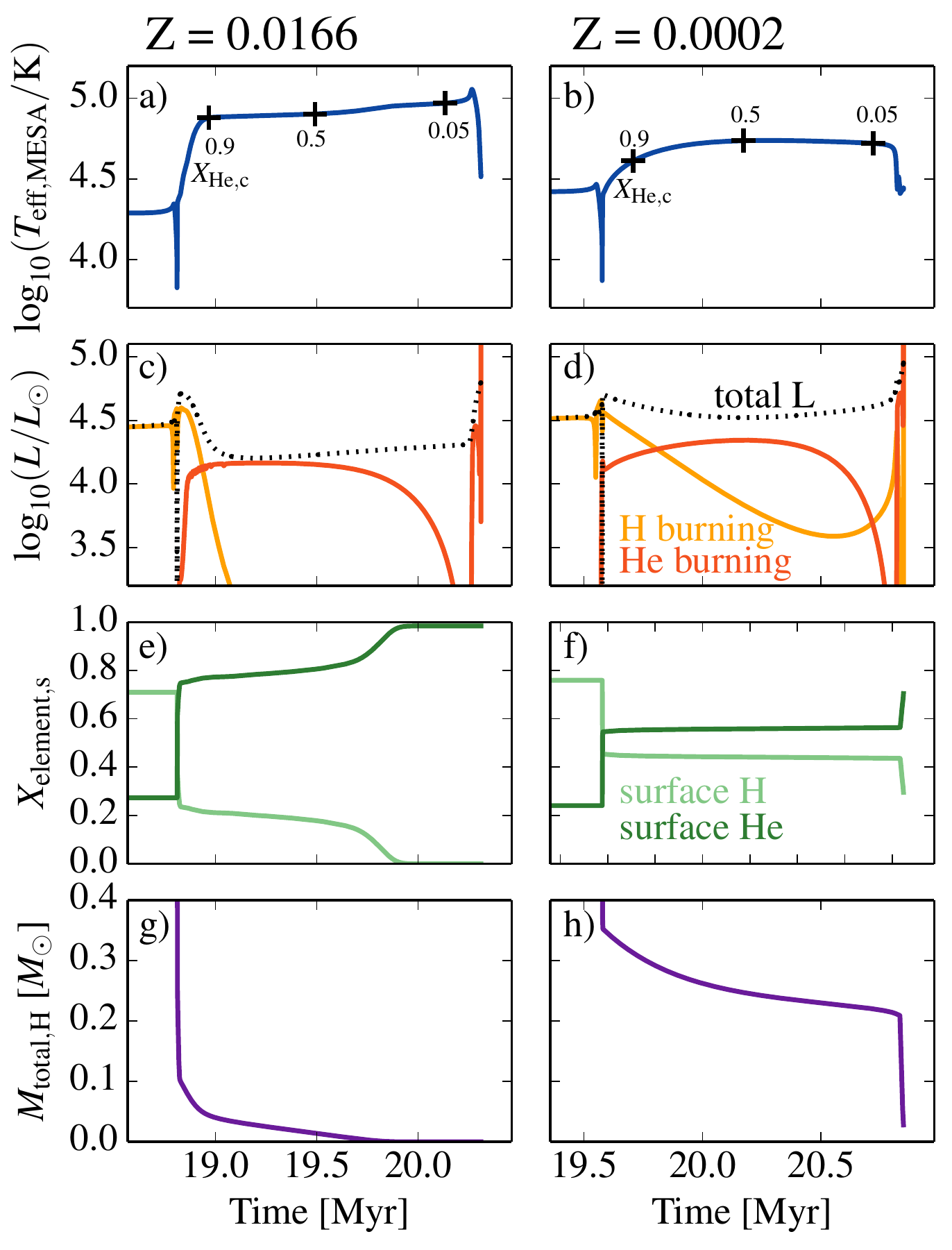}
\caption{We show the effective temperature (panels a and b), luminosity from hydrogen and helium burning (panels c and d),  surface mass fraction of hydrogen and helium (panels e and f) and total mass of hydrogen (panels g and h) for our $Z = 0.0166$ (left column) and $Z = 0.0002$ (right column) models from slightly before \lang{Roche-lobe} overflow and for the rest of the evolution. In the panels with effective temperature \lang{plus signs indicate} the times at which the central helium mass fraction has reached 0.9, 0.5 and 0.05. In the luminosity panels we show the total stellar luminosity with a black, dotted line.}
\label{fig:evol_strip}
\end{figure}

\subsection{The resulting stripped star: \lang{The} long-lived helium burning phase and further evolution}\label{sec:evol_Z_strip}

In the previous section we described that the stripping process is inefficient at lower metallicity: it fails to remove entire envelope. Our metal-poor models consist of a helium core surrounded by a remaining layer of envelope material of more than 1\Msun (right panel of \figref{fig:Kipp_M12}). This layer \lang{consists} of a mixture of hydrogen and helium, containing a total amount 0.38\Msun of pure hydrogen right after detachment from the Roche lobe. For our metal-rich model the remaining shell contains less than 0.05\Msun of hydrogen, which initially allows for weak burning in a shell around the core (left panel of \figref{fig:Kipp_M12}). When investigating even lower metallicities ($Z \leq 0.00002$), we find that the star never becomes hotter than its zero-age main sequence because of the large amount of leftover hydrogen after mass transfer.

This has consequences for the luminosity and effective temperature. The metal-poor stripped stars are about 30~000~K cooler\lang{, 0.3 dex brighter,} and almost four times bigger than their metal-rich counterparts\lang{;} see \figref{fig:MESA_hestar_prop} and \tabref{tab:hestar_prop}. This is somewhat counterintuitive, since it is the opposite of what is found for single and pre-interaction stars. Before interaction, we find metal-poor stars to be more compact and hotter. After stripping we find metal-rich stars to be more compact and hotter.   

The difference in mass, luminosity and composition affects the remaining lifetime. Panel b in \figref{fig:MESA_hestar_prop} shows that the remaining lifetime increases with metallicity by about 20\% (see panel b). This is \lang{because of} the lower mass of the stripped stars at higher metallicity. The fact that our metal-poor stars still provide part of their luminosity through H burning in a shell \lang{does not prolong} their life.   

The further panels in \figref{fig:MESA_hestar_prop} provide an overview of various properties of stripped stars, when they are halfway their helium burning phase (defined as the moment when $X_{\text{He,c}} = 0.5$). The shaded bands spans the variation in properties during the hot phase of their helium burning lifetime (which we define as $ 0.9 > X_{\text{He,c}} > 0.05$). 

In \figref{fig:evol_strip} we compare the evolution as a function of time during the helium burning phase of our low metallicity model ($Z=0.0002$) and our \lang{metal-rich} reference model. In the top row we compare the evolution of the effective temperature as a function of time. Plus symbols \lang{indicate} the central helium mass fractions $X_{\text{He,c}} = 0.9, 0.5$ and 0.05 \lang{during core helium burning}. For our metal-rich model, we see the first quick initial rise of the effective temperature, when the star is still contracting within its Roche lobe. \lang{Afterward}, the effective temperature steadily rises by about 0.1 dex over the course of the helium burning phase. At the end of the helium burning phase\lang{,} the temperature rises \lang{briefly until} helium shell burning ignites and the star expands\lang{; this is} visible in the diagram as a drop in the effective temperature. In contrast, our metal-poor \lang{model remains} substantially cooler for the first part of the helium burning phase. The temperature slowly rises and \lang{settles at about $\log _{10} (T_{\text{eff}, \textsc{mesa}}/\text{K}) = 4.7$ only after a third of the helium burning phases}. 

In the second row in \figref{fig:evol_strip} we show the luminosity produced by hydrogen and helium burning separately over time for our metal-poor and metal-rich model. Shortly after the end of Roche-lobe overflow, we see the helium burning luminosity quickly rises but hydrogen burning \lang{still provides} most of the energy and exceeds the helium burning luminosity about 0.5 dex. As the star evolves along the helium burning main sequence, we see that the contribution of hydrogen burning quickly drops in the metal-rich model, as expected from our earlier discussion of the very thin shell that remains and the effect of the stellar winds. In contrast, the thick H-rich shell present in the metal-poor case remains actively burning hydrogen throughout the full helium burning phase. It does however weaken and after about 20\% of the helium burning lifetime, helium burning takes over as the dominant source of energy. This is roughly around the same time \lang{at which} the effective temperature of the star stabilizes, as can be seen in the panel above. 

In the third row of \figref{fig:evol_strip} we compare the evolution of the H and He surface mass fraction as a function of time. At the far left of the diagram, we see the quick reversal of H and He due to \lang{Roche-lobe} stripping. In both cases He is the dominant element at the surface. In the case of the metal-rich star we see the effect of stellar winds slowly removing the outer layer containing H. After about \lang{two-thirds} of the helium burning phase\lang{,} the winds have removed the last remaining H and the surface abundance of H quickly drops to zero. We find this behavior in all our models with metallicity $Z \geqslant 0.01$.  

In contrast, for the metal-poor model we see that the surface abundance of H and He are constant throughout the helium burning phase, since \lang{the wind mass loss} is negligible. At the end of the evolution, during helium shell burning we find that the star fills its Roche lobe again. This removes part of the \lang{H-rich} layer, but not all \lang{of this layer}. At metallicities $Z \leqslant 0.0047$, the stripped stars have enough hydrogen-rich envelope left at this stage to fill the Roche lobe a second time. Because the remaining evolution is very rapid, these stars are likely to end their lives during mass transfer. 

In the last row of \figref{fig:evol_strip} we show the total mass of hydrogen present in the star. In the metal-rich case, the star is deeply stripped during Roche-lobe overflow. During the helium core burning phase the star loses more mass through stellar winds. As a result, the total hydrogen mass is $M_{\text{H,tot}} \sim 0.05$~\Msun right after the end of Roche-lobe overflow, but completely disappears before explosion \lang{owing to mass loss by winds}. The metal-rich stripped star is expected to give rise to a type Ib supernova. 

In the metal-poor case, the Roche-lobe overflow phase leaves $M_{\text{H,tot}} \sim 0.35$~\Msun. Mass loss by winds is negligible in this case, but shell burning significantly decreases the amount of hydrogen by converting it into helium. At the end of the helium burning phase $\sim 0.2$ is left. During helium shell burning the stripped star swells up and fills its Roche lobe a second time. The second phase of mass transfer decreases the total hydrogen mass to $\sim 0.03$~\Msun at the end of our calculations. When the metal-poor stripped star ends its life, it \lang{still thus shows} signatures of hydrogen in the very early spectra of the supernova.

\begin{figure*} 
\centering
\includegraphics[width=0.9\hsize]{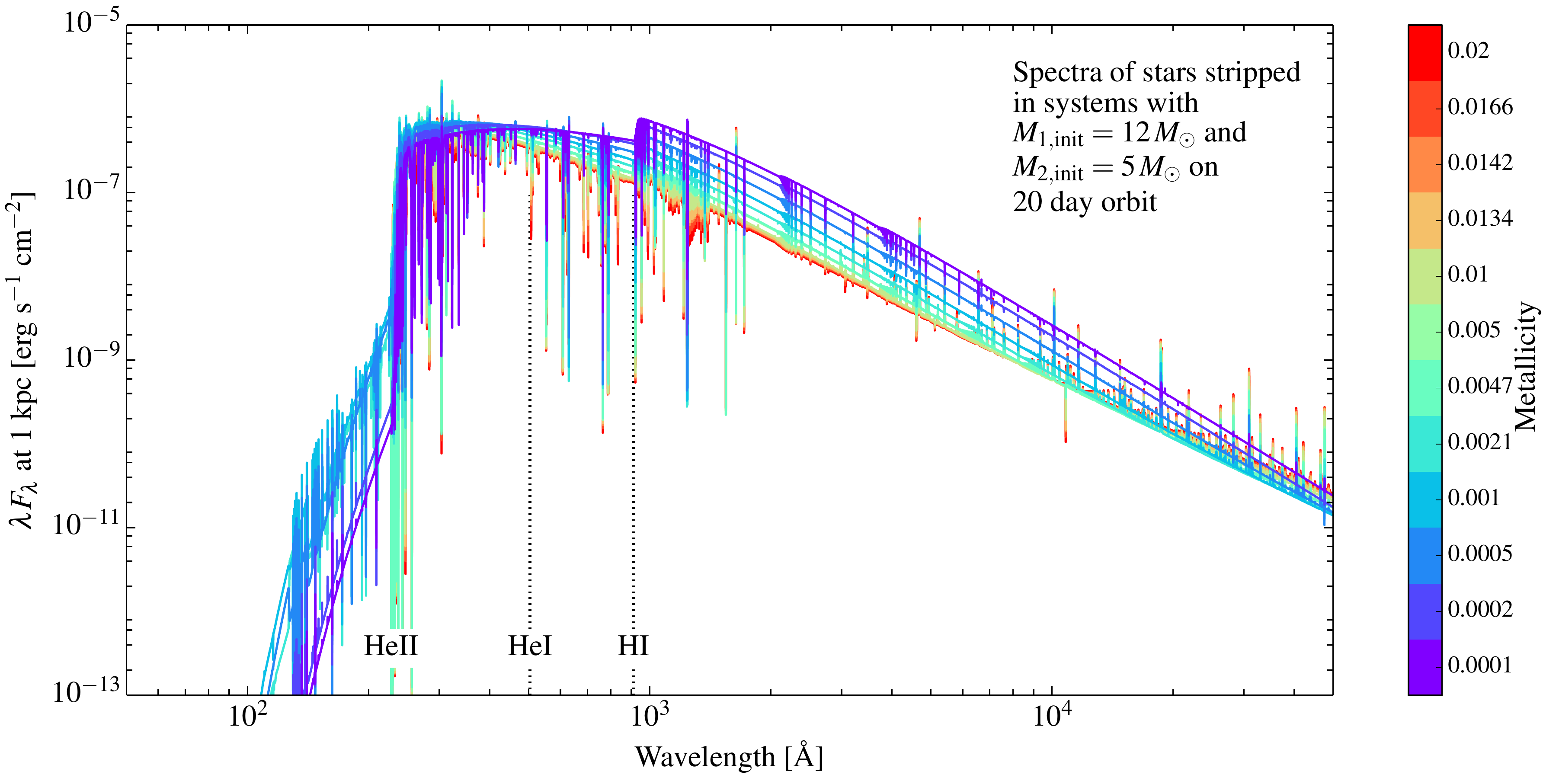}
\caption{Spectral energy distributions for stripped stars resulting from a 12\Msun donor star for a range of different metallicities. These are modeled with CMFGEN assuming our standard wind mass loss rates. We \lang{indicate} the ionization limits of hydrogen and helium with dotted lines. At high metallicity, the mass loss rates are higher, leading to a denser wind surrounding the star. Recombination in the wind of He$^{2+}$ to He$^{+}$ blocks the emission of photons with wavelengths shorter than $\lambda = 228\,$\AA\ for models with $Z \geqslant 0.0047$. The spectra of the metal-rich models show many characteristic emission lines. At lower metallicity, the winds are transparent and most lines are in absorption. The metal-poor stripped stars still have hydrogen at their surface. This is responsible for the characteristic drop at 912\,\AA. \lang{The stripped stars} are also more luminous and somewhat cooler. All model spectra are created during core helium burning when central helium abundance is $X_{\text{He, c}} = 0.5$.}
\label{fig:SED_Z}
\end{figure*}

\section{Effect of metallicity (II): \lang{The} emerging spectra of stripped stars}\label{sec:CMFGEN_Z}

In the previous section, we showed that stripped stars at higher metallicity have hotter surfaces, are less luminous, have stronger stellar winds\lang{,} and contain less hydrogen at their surface. Here, we discuss how the combined effects of metallicity influence the spectral energy distribution and the characteristic spectral features. For this we use our CMFGEN atmosphere simulations created for the stripped stars discussed in the previous section, taken when they are halfway through their central helium burning phase ($X_{\rm He, c} = 0.5$).

\begin{figure*}
\centering
\includegraphics[width=.32\textwidth]{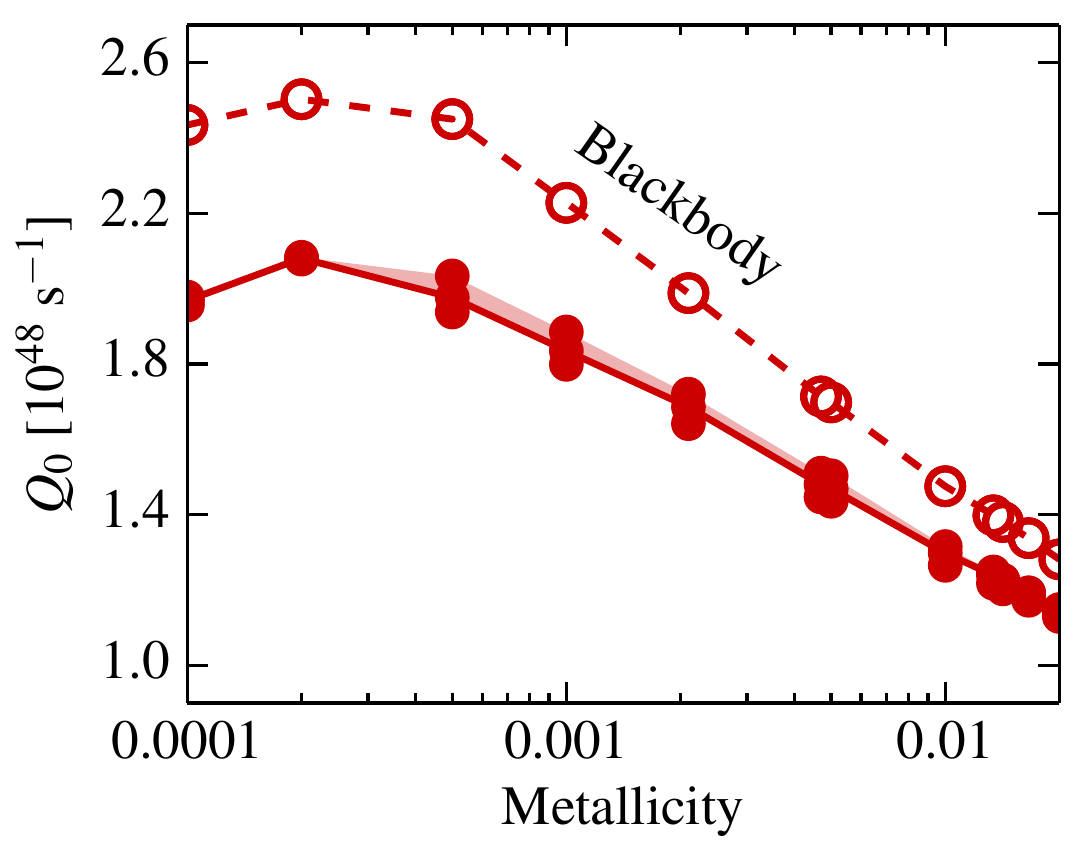}
\includegraphics[width=.32\textwidth]{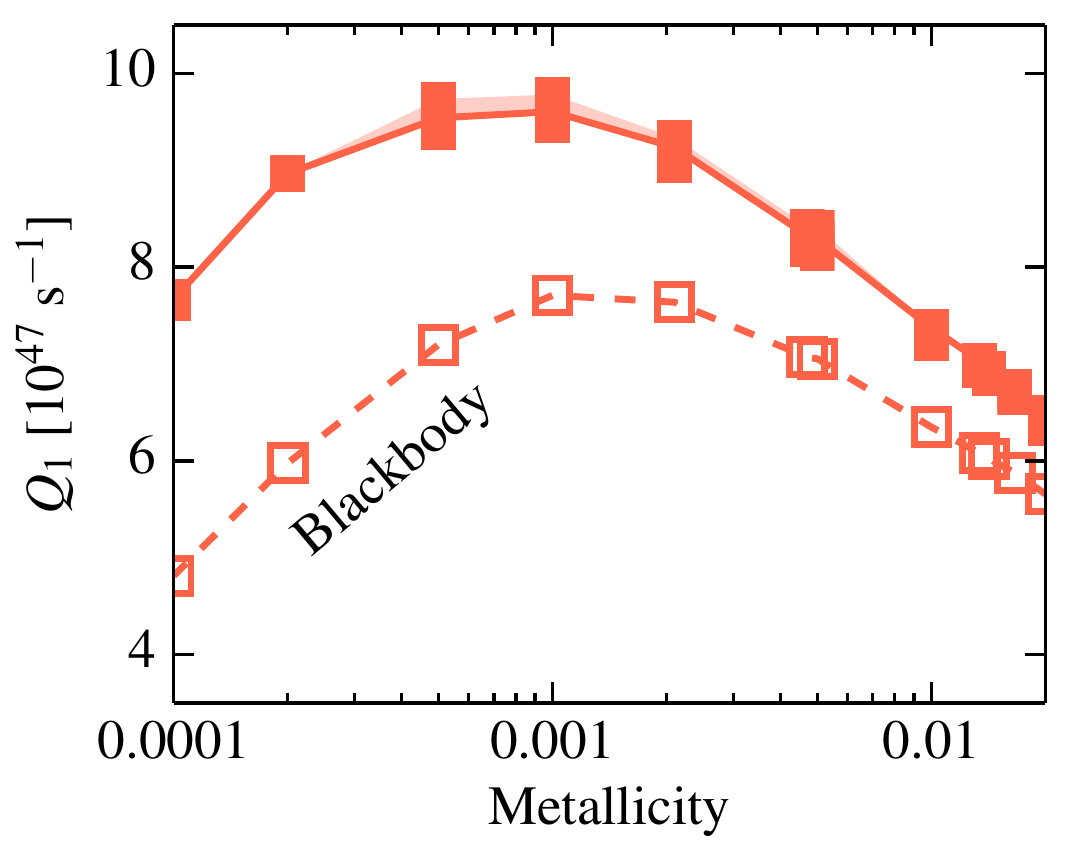}
\includegraphics[width=.33\textwidth]{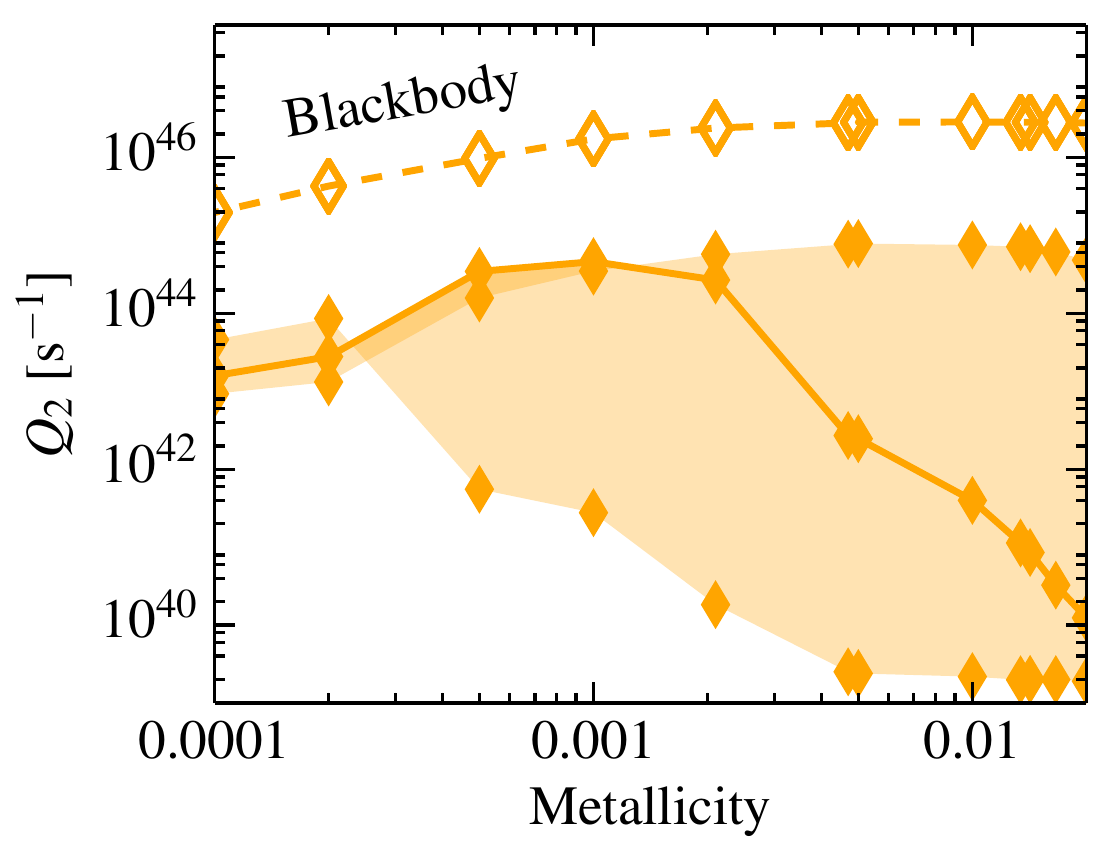}
\caption{Flux of HI ($Q_0$, red circles), HeI ($Q_1$, orange squares) and HeII ($Q_2$, yellow diamonds) ionizing photons from our models of stripped stars with various metallicity. The flux \lang{estimates} using the effective temperature of the evolutionary models for blackbody radiation \lang{are} shown with open symbols and dashed lines. The more accurate \lang{estimates} from the modeled spectra are shown with filled symbols and solid lines. The wind mass loss rate affects the appearance of the spectral energy distribution (see \figref{fig:SED_Z}) and \lang{thus the} emission rate of ionizing photons \lang{as well}. We show the flux estimates between \lang{the mass loss rate that is} three times higher ("enhanced") and three times lower ("reduced") with shaded regions. Both $Q_0$ and $Q_1$ are plotted on linear scale as the variation is small with metallicity. \lang{The value} $Q_2$ is strongly wind mass loss dependent and therefore varies significantly with metallicity.}
\label{fig:Q_Z}
\end{figure*}

\subsection{Spectral energy distribution \lang{and} flux of ionizing photons}
 
 In \figref{fig:SED_Z} we show the spectra in conventional units ($\lambda F_{\lambda}$ at 1 kpc in erg\,s$^{-1}$\,cm$^{-2}$ versus $\lambda$ in \AA, where $\lambda$ is the wavelength and $F_{\lambda}$ the flux emitted at that wavelength) for stripped stars with metallicities between $Z = 0.0001$ and $0.02$. The flux of all spectra peak in the extreme \lang{UV at} wavelengths between the thresholds for ionization of HeI and HeII (\lang{indicated} with vertical dashed lines). The metal-rich models peak at slightly shorter wavelengths. This is due to their higher surface temperatures. The metal-rich models also have stronger and denser winds. In principle, this can move the photosphere outward to larger radii, resulting in a reduction of the effective temperature that characterizes the emerging spectrum. However, for our standard mass loss rates we find that the winds are transparent in all cases. The radius of the stellar surface, $R_{\star}$, and the photosphere,  $R_{\text{eff}}$, coincide\lang{;} cf.\ \tabref{tab:hestar_prop}. The metal-poor models are brighter and show the distinctive trough at 912\,\AA\, at the threshold for hydrogen ionization. This feature is weaker in the metal-rich models. They still have traces of hydrogen at their surface abundances ($X_{\rm H, s} \sim 0.2$, but this is completely ionized\lang{)}.  

Our spectral models show that  stripped stars are very efficient emitters of ionizing photons at all metallicities. In our high metallicity reference model we find that \lang{the stripped star} emits 85\% of their energy as HI ionizing photons, 60\% as HeI ionizing photons\lang{,} and $5 \times 10^{-6}$~\% as HeII ionizing photons. In \figref{fig:Q_Z} we show the metallicity dependence of the total number of ionizing photons emitted per second, $Q_0$, $Q_1$, and $Q_2$ at wavelengths shorter than the ionization potential for HI, HeI and HeII, respectively\lang{;} see also \tabref{tab:Q_Z}. We show the results for the three variations adopted for the stellar wind mass loss rates: standard, three times enhanced\lang{,} and three times reduced. For reference, we also show an estimate for a blackbody spectrum with the same temperature as the stellar surface, i.e.\ the effective temperature given by the MESA models (open symbols). 

\subsubsection*{HI ionizing photons }

The flux of HI ionizing photons (left panel of \figref{fig:Q_Z}) is about  $10^{48} $ s$^{-1}$ and rises by about a factor two with decreasing metallicity. This can be understood as the result of two effects that counteract each other. As we have shown in \secref{sec:evol_Z_rlof}, the metal-poor models are not completely stripped by Roche-lobe overflow; a hydrogen-rich layer is left at the surface. This results \lang{in} stripped stars that are more massive, luminous and slightly cooler (see \figref{fig:MESA_hestar_prop}). The higher luminosity favors the production of ionizing photons, while the lower temperatures disfavor it. The net effect is a mild increase of ionizing photons with decreasing \lang{metallicity with} a peak at $Z=0.0002$. 

We only find a variation of $\lesssim 2\%$  in $Q_0$ when varying the assumed mass loss rate. Our results are thus robust against uncertainties in the mass loss rate.

A simple blackbody estimate for hydrogen ionizing flux (open symbols)  based on the surface temperature given by MESA is remarkably accurate. It overestimates $Q_0$ only by about 10\%. This is potentially interesting, since the blackbody approximation provides a computationally cheap alternative for the detailed atmosphere simulations that \lang{we conducted}. The reason for the small difference between the detailed simulations and the blackbody estimate is that hydrogen is almost completely ionized throughout the wind (see \figref{fig:ionisation_structure_M12}). Without neutral hydrogen present, the hydrogen ionizing photons \lang{cannot} be used for hydrogen ionization in the wind and propagate into the surroundings. 

\subsubsection*{HeI ionizing photons}

We find very similar trends for the flux of HeI ionizing photons (central panel of \figref{fig:Q_Z}), which is only a factor of two below the HI ionizing flux. The HeI ionizing flux peaks at a metallicity $Z = 0.001$ and also closely follows the estimate from a simple blackbody. This is because neutral helium is also almost completely depleted in the wind. The detailed spectral simulations give a 20\% higher flux than the simple blackbody estimate. The predictions for the HeI ionizing flux are also robust against uncertainties in the mass loss rates. 

\subsubsection*{HeII ionizing photons}

The flux of HeII ionizing photons (right panel of \figref{fig:Q_Z}) is strongly reduced and is 4-8 orders of magnitude below the estimates for the HI and HeI ionizing photon flux \lang{(the scale is different on the vertical axis in the three different panels)}. The results are extremely sensitive to the assumed mass loss rate (shaded band), giving variations of almost 6 orders of magnitude at high metallicity. This is because the helium recombination front is very sensitive to the assumed mass loss rate \lang{and therefore} we cannot accurately estimate the emission rate of HeII ionizing photons. \lang{We} find that the sensitivity our multiplicative variations in the mass loss rate are \lang{only reduced for very low metallicity ($Z \leq 0.0002$) where the stellar winds become insignificant}, but still leads to variations of an order of magnitude. 

A blackbody estimate for HeII ionizing flux is not appropriate. It overestimates the flux by about two orders of magnitude at least. The reason is that helium is not completely ionized throughout the wind and in the outer parts the density is high enough and temperature low enough for He$^{2+}$ to recombine to He$^{+}$. This leads to \lang{high-energy photons that are} used for ionizing the wind instead of emerging to the surroundings. We conclude that the HeII ionizing photons are too uncertain at present to provide meaningful quantitative predictions. 

\begin{table*}
\centering
\caption{Ionizing flux emitted from our models of stripped stars. The columns 2, 3\lang{,} and 4 show the fraction of the emitted luminosity that is HI, HeI\lang{,} and HeII ionizing\lang{,} respectively. The last three columns show the number of ionizing photons emitted per second from each model. In parenthesis we show the values for the reduced and enhanced mass loss rate models of the given metallicity.}
\label{tab:Q_Z}
\small
\begin{tabular}{lccccccccc}
\toprule\midrule
$Z$ & $L_{\text{HI}}/L_{\text{tot}}$ & $L_{\text{HeI}}/L_{\text{tot}}$ & $L_{\text{HeII}}/L_{\text{tot}}$ & \multicolumn{2}{c}{$\log_{10} Q_0$\_st (red-enh)} & \multicolumn{2}{c}{$\log_{10} Q_1$\_st (red-enh)} & \multicolumn{2}{c}{$\log_{10} Q_2$\_st (red-enh)}\\ 
 &  &  &  & \multicolumn{2}{c}{s$^{-1}$} & \multicolumn{2}{c}{s$^{-1}$} & \multicolumn{2}{c}{s$^{-1}$}\\ 
\midrule
0.02 & 0.853 & 0.602 & $1.8\times 10^{-8}$ & 48.1 & (48.1-48.1) & 47.8 & (47.8-47.8) & 40.1 & (44.7-39.3)\\ 
\rowcolor{black!10}0.0166 & 0.854 & 0.604 & $4.6\times 10^{-8}$ & 48.1 & (48.1-48.1) & 47.8 & (47.8-47.8) & 40.5 & (44.8-39.3)\\ 
0.0142 & 0.854 & 0.605 & $1.2\times 10^{-7}$ & 48.1 & (48.1-48.1) & 47.8 & (47.8-47.8) & 40.9 & (44.8-39.3)\\ 
0.0134 & 0.853 & 0.604 & $1.5\times 10^{-7}$ & 48.1 & (48.1-48.1) & 47.8 & (47.8-47.8) & 41.1 & (44.9-39.3)\\ 
0.01 & 0.853 & 0.605 & $5.1\times 10^{-7}$ & 48.1 & (48.1-48.1) & 47.9 & (47.9-47.9) & 41.6 & (44.9-39.3)\\ 
0.005 & 0.847 & 0.599 & $2.8\times 10^{-6}$ & 48.2 & (48.2-48.2) & 47.9 & (47.9-47.9) & 42.4 & (44.9-39.4)\\ 
0.0047 & 0.846 & 0.597 & $3.0\times 10^{-6}$ & 48.2 & (48.2-48.2) & 47.9 & (47.9-47.9) & 42.4 & (44.9-39.4)\\ 
0.0021 & 0.828 & 0.573 & $2.8\times 10^{-4}$ & 48.2 & (48.2-48.2) & 48.0 & (48.0-48.0) & 44.4 & (44.8-40.3)\\ 
0.001 & 0.798 & 0.535 & $4.3\times 10^{-4}$ & 48.3 & (48.3-48.3) & 48.0 & (48.0-48.0) & 44.7 & (44.5-41.4)\\ 
0.0005 & 0.746 & 0.471 & $2.9\times 10^{-4}$ & 48.3 & (48.3-48.3) & 48.0 & (48.0-48.0) & 44.5 & (44.2-41.7)\\ 
0.0002 & 0.681 & 0.395 & $2.1\times 10^{-5}$ & 48.3 & (48.3-48.3) & 48.0 & (48.0-48.0) & 43.4 & (43.1-43.9)\\ 
0.0001 & 0.613 & 0.329 & $1.2\times 10^{-5}$ & 48.3 & (48.3-48.3) & 47.9 & (47.9-47.9) & 43.2 & (43.0-43.7)\\ 
\bottomrule
\end{tabular}

\tablefoot{Our reference models are marked with \lang{gray} background.}
\end{table*}

\begin{figure*}
\centering
\includegraphics[width=\hsize]{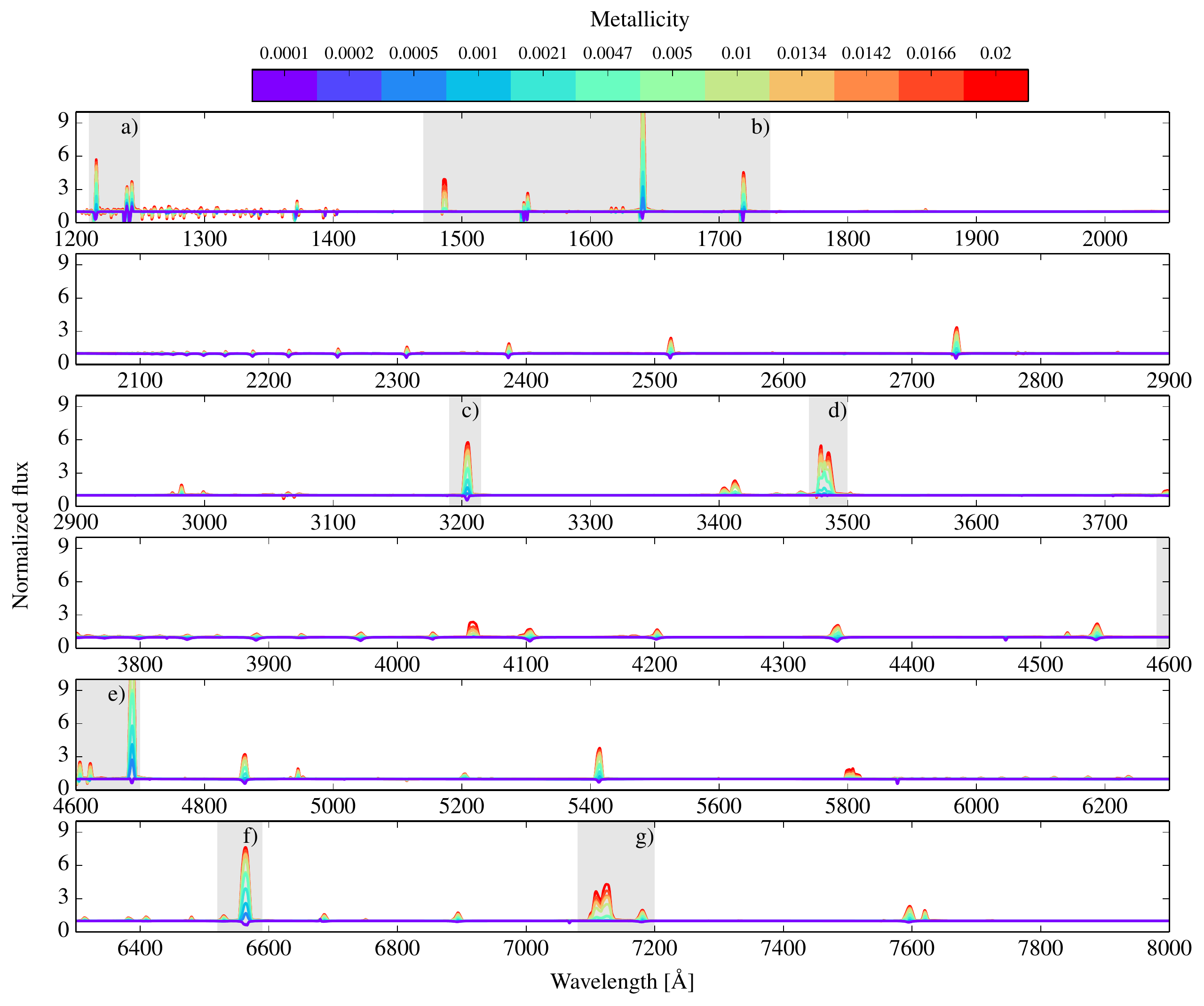}
\includegraphics[width=\hsize]{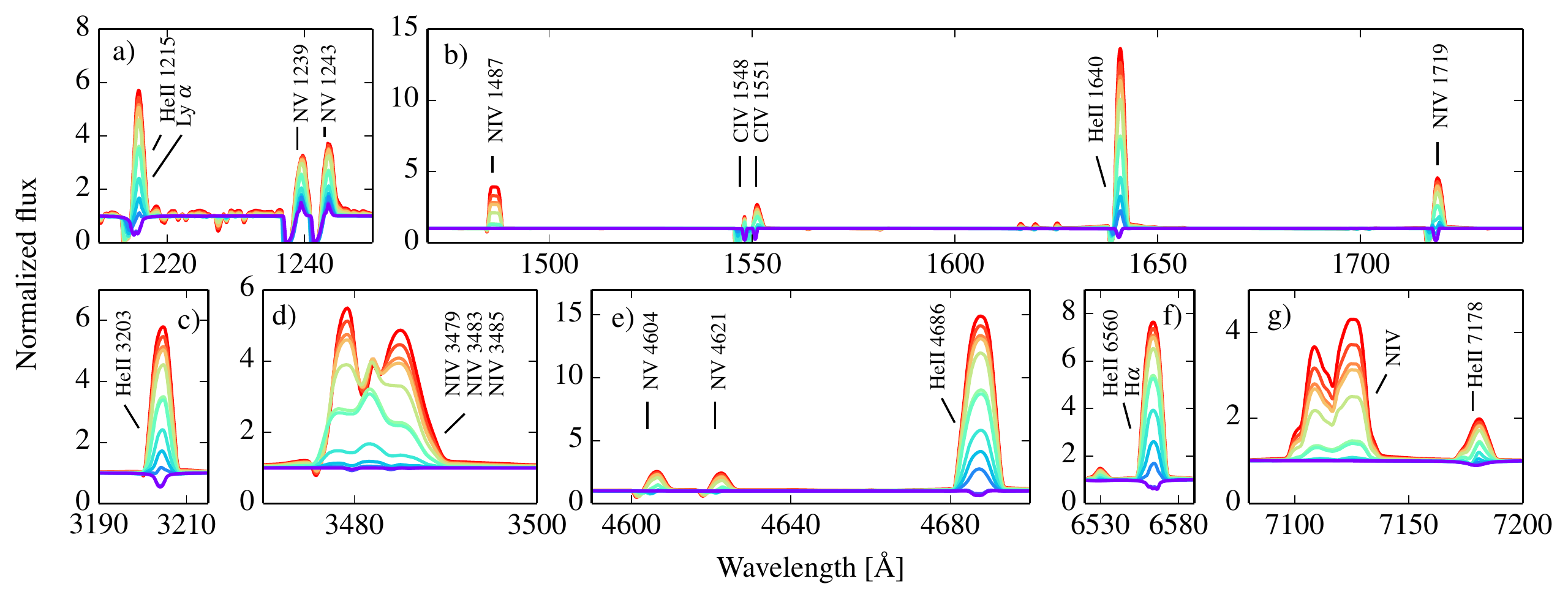}
\caption{Spectra of the atmosphere models with different metallicities and the standard mass loss rates. At lower metallicities the line strengths decrease until they even go into absorption. In the bottom panel we zoom in on a few regions with strong lines. One of the strongest lines is the HeII~$\lambda$4686 line which also goes into absorption when $Z \leqslant 0.0002$.}
\label{fig:spectra_Z}
\end{figure*}

\subsection{Spectral features}

The spectra of metal-rich stripped stars show a rich forest of emission lines spanning not only the extreme UV where the emission peaks, but also the near UV, optical and near infrared which are accessible by \lang{ground-based} facilities. 

At lower metallicity the stripped stars have lower mass loss rate and lower surface temperatures, which has implications for the spectral signatures. The emission lines decrease in strength with decreasing metallicity. This can be seen most clearly in the \figref{fig:spectra_Z}, which shows the normalized spectra with inset panels that zoom in on the most prominent lines. Most of these lines are recombination lines (see e.g.\ panels a, c and e in \figref{fig:spectra_Z}). The strong lines we pointed out for our reference model in \secref{sec:M12_CMFGEN} remain the most important lines, but with lower metallicity they decrease in strength and at $Z \leqslant 0.0002$ they turn into absorption lines. We also see changes in the shape of the line. We find a sequence changing from absorption lines into P~Cygni profiles. The Ly$\alpha$ and HeII~$\lambda$1215 blend in panel a of \figref{fig:spectra_Z} is an example. In other cases we find absorption lines changing into emission profiles with increasing mass loss rate\lang{;} see for example the HeII~$\lambda$4686.

\section{Discussion \lang{and i}mplications}\label{sec:discussion}

A full assessment of the implications will require larger model grids spanning different initial masses, periods, mass ratios and a more extensive exploration of the uncertainties. However, based on the insight resulting from the exploratory calculations presented here, we foresee several implications. We discuss them briefly in this section and speculate on the basis of the very simple estimates that we can make at present. 

\subsection{Budget of ionizing photons and implications for cosmic reionization}

Quantifying the budget of ionizing photons produced by stellar populations is of wide interest for a variety of applications. These range from cosmological simulations that assess the reionization of the intergalactic medium (IGM), spectral synthesis simulations used to understand the properties of the strong emission lines of galaxies at intermediate to high redshift and HII regions nearby.

Massive Wolf-Rayet (WR) stars are the stars in stellar populations that emit ionizing photons with the highest rate. We explored the properties of stripped stars that are only produced in binaries. We showed that these stripped stars exhibit high effective temperatures for all metallicities we considered. They emit the majority of their flux as HI and HeI ionizing photons. 

To provide a first ballpark estimate of the relative emission rate of stripped stars, we compare our models with a typical WR star. Stripped stars are the \lang{lower mass} counterpart of WR stars as they are also the stripped helium core of a massive star. \lang{Wolf-Rayet} stars are more luminous compared to stars stripped in binaries. However, \lang{WR stars} have shorter lifetimes and they are not favored by the stellar initial mass function. For the typical values for a \lang{WR} star\lang{,} we \lang{used} the WC model by \citet[][stage 48]{2014A&A...564A..30G}, which corresponds to an initial mass $M_{\text{WR, init}} = 60\,\Msun$ single star. This star spends about $\Delta t_{\text{WR}} \sim 0.4\,\text{Myr}$ in the WR phase, during which it emits $Q_{0, \text{WR}} \sim 2.8\times10^{49}$ HI ionizing photons per second. We \lang{assumed} all stars with this initial mass become WR stars at some point during their lives ($f_{\text{WR, }60\,M_{\odot}} = 1$).  

For stripped stars we take our reference model with standard mass loss rate, which has an initial mass of 12\Msun. We \lang{assumed} that a fraction $f_\text{strip} = 0.33$ of stars with this initial mass becomes stripped \citep{2012Sci...337..444S}. Using a \citet{1955ApJ...121..161S} initial mass function, the relative contribution can then be estimated as
\begin{equation}\label{eq:f_est}
\begin{split}
\eta &= \dfrac{f_{\text{strip}}}{f_{\text{WR, }60\,M_{\odot}}} \times \dfrac{Q_{0,\text{ strip}}}{Q_{0, \text{WR}}} \times \dfrac{\Delta t_{\text{strip}}}{\Delta t_{\text{WR}}} \times \left( \dfrac{M_{\text{strip, init}}}{M_{\text{WR, init}}} \right) ^{-2.35} \\ &= \dfrac{0.33}{1.0} \times \dfrac{1.19 \times 10^{48}\, \text{s}^{-1}}{2.8 \times 10^{49}\, \text{s}^{-1}} \times \dfrac{1.2\, \text{Myr}}{0.4\, \text{Myr}} \times \left( \dfrac{12\, M_{\odot}}{60\,M_{\odot}} \right) ^{-2.35} \approx 1.8
\end{split}
\end{equation}
\noindent With this very simple estimate, we find that the stripped stars produce almost twice the amount of ionizing photons in comparison to the WR stars. More extensive modeling is needed to assess the full contribution for a realistic population, but it seems likely \lang{that} stripped stars make a significant contribution to the total budget of ionizing photons. 

An accurate estimate would require extensive spectral model grids of stripped stars spanning over mass. This is beyond the scope of the paper, but a full grid will be presented in G\"{o}tberg et al.\ (in prep.). However, a boost of only a factor of a few would, in principle, be enough to complete cosmic reionization by redshift 6--7 \citep[see][for a review]{2014ARA&A..52..415M}.

The most promising aspect of stripped stars, as potential contributors of the photons needed for reionization, is that they emit them with a time delay. The progenitor star first has to complete its main-sequence evolution, which takes about 20\,Myr for the model presented here. Allowing for different progenitors with different masses and lifetimes we expect that the boost of ionizing photons comes with delay times ranging from a few to at least 100 \,Myr or more (G\"otberg et al. in prep., see also predictions by \citealt{2016MNRAS.456..485S}).  Observations of nearby star clusters of this age suggest that this is sufficient time to remove most of the remaining gas \citep[e.g.][]{2010A&A...517A..50G, 2014MNRAS.445..378B}. Numerical simulations also indicate that large local and temporal variations of the escape fractions are possible as feedback of massive stars removes gas and clears lines  \citep[e.g.][]{2011A&A...530A..87P, 2015MNRAS.453..960M, 2017arXiv170500941T}. 

This suggests that the slightly delayed photons produced by stripped stars have a much larger chance to escape and become available to ionize the intergalactic medium. Quantitatively assessing the impact of this requires reliable estimates of the escape fraction, which are not available at present. A boost of an order of magnitude in the escape fraction of ionizing photons for stripped stars does not appear \lang{to be unreasonable in} light of the findings in the high resolution simulations presented by \citet{2017arXiv170500941T}\lang{,} who report temporal variations in the escape fraction that reach up \lang{to six} orders of magnitude.

Finally, we note briefly that binary stars have other ways to change the spectral energy distributions of stellar populations, apart from producing stripped stars as we discussed here. Substantial contributions are expected from mass gainers and stellar mergers, which effectively repopulate the upper end of the initial mass function as blue stragglers \citep[e.g.][]{2014ApJ...782....7D, 2015ApJ...805...20S, 2014ApJ...780..117S}. The BPASS simulations suggest that these also make an important contribution \citep[e.g.][]{2016MNRAS.456..485S}. 

Furthermore, a subset of \lang{binaries remain} bound after the first star ends its life as a neutron star or black hole. A fraction of \lang{these evolve} through an X-ray binary phase. A possible importance of their contribution to heating and reionization of the intergalactic medium (IGM) at high redshift has also been considered by various authors \citep[e.g.][]{2011A&A...528A.149M, 2013ApJ...764...41F, 2017ApJ...840...39M}. However, \citet{2017ApJ...840...39M} \lang{have concluded} that the contribution of X-ray binaries to the ionization of the bulk IGM is negligible.

\begin{figure*}
\centering
\includegraphics[width=0.65\hsize ,trim={0 0 0 3cm},clip]{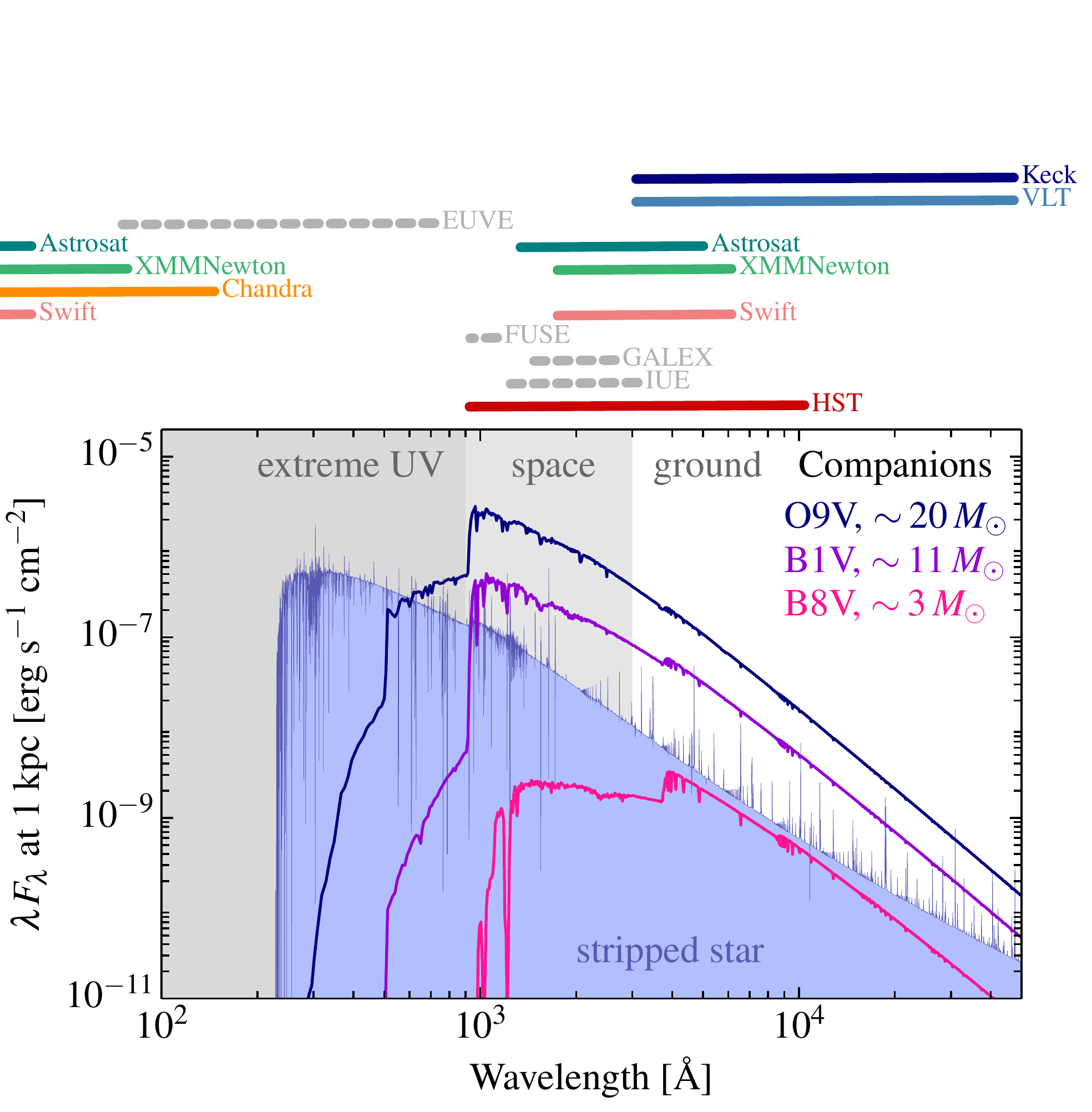}
\caption{\lang{Our} standard mass loss reference model of a stripped star in shaded blue area together with three Kurucz models corresponding to main sequence companion stars of spectral types O9V, B1V\lang{,} and B8V (approximate masses 20, 11\lang{,} and 3\,$M_{\odot}$, parameters given in \tabref{tab:Kurucz}). \lang{The} available wavelength bands from the ground, from space\lang{,} and in the extreme UV \lang{are shaded in white and gray backgrounds}. Above the plot we show the wavelength ranges probed by various telescopes\lang{; colors show} active \lang{telescopes} and gray \lang{shows} deactivated telescopes. In the extreme UV where the bulk of the emission from the stripped star is dominant, no telescope is currently available. In the optical there are many available instruments, but in this region \lang{even low-mass} companions could clearly dominate the emission of the binary star. With \lang{lower mass} companions some of the strong emission lines of the stripped star could pierce through the spectrum of the companion, but with higher mass companions it would be very difficult to see a stripped companion from optical spectral analysis.}
\label{fig:SED_comp}
\end{figure*}

\begin{table}
\centering
\caption{Kurucz models used in \figref{fig:SED_comp} to illustrate spectra of companion stars taken from \cite{2011MNRAS.413.1515H}. }
\label{tab:Kurucz}
\small
\begin{tabular}{lccccc}
\toprule\midrule
Kurucz model & SpT & $M$ & $L$ & $T_{\mathrm{eff}}$ & $\log_{10} g$ \\  
 &  & [$M_{\odot}$] & [$L_{\odot}$] & [K] & [cm/s$^{-2}$] \\  
\midrule

t33000g40 & O9V & 20 & 60039 & 33000 & 4  \\ 
t25000g40 & B1V & 11 & 10665 & 25000 & 4  \\ 
t12000g40 & B8V & 3 & 153 & 12000 & 4  \\ 
\bottomrule 
\end{tabular}
\end{table}

\subsection{Observability of stripped stars and strategies \lang{to find these stars}}

Despite the expectation that a large fraction of massive \lang{stars interact} in a binary system during their lifetime and produce a long-lived stripped star, only very few systems have been detected so far \citep[see][]{2008A&A...485..245G, abels_paper}. This apparent paradox may simply be the result of biases and selection effects \citep[e.g][]{2014ApJ...782....7D}, but this has not been quantified. Our simulations provide insight \lang{into} these biases and can guide efficient observing campaigns that aim to increase the observed sample of stripped stars.

The main challenge is that hot, stripped stars do not appear in isolation. They still reside in orbit around their companion star, which \lang{typically dominates} the optical and UV flux \lang{heavily}. This is illustrated in \figref{fig:SED_comp}, where we show the SED of our reference model (blue shading) together with the spectra of three possible main sequence companion stars, roughly spanning the range of companions that may be expected after interaction. The companion spectra come from the ATLAS9 models \citep{2004astro.ph..5087C,2011MNRAS.413.1515H}\lang{;} see \tabref{tab:Kurucz} for an overview of the adopted parameters. \lang{The companion models} roughly correspond to evolutionary tracks of a 20, 11, and 3 \Msun star halfway central hydrogen burning (defined as $X_{H,c} = 0.5$). Their properties are still close to their zero-age main sequence properties, as we expect for relatively \lang{unevolved} or rejuvenated companions.

One possible strategy is to detect stripped stars through the UV excess that would be detected in otherwise apparently normal, main sequence stars. However, as \figref{fig:SED_comp} shows, the emission of stripped stars peaks in the extreme UV, which is not accessible from the ground \lang{or} space with present day observing facilities \footnote{\lang{The} only mission that systematically explored the extreme UV is the all-sky Extreme Ultra Violet Explorer (EUVE; operational in 1992--2001)}. The UV regime between $912-3200$\,\AA\, is accessible to present-day facilities or can be mined in the FUSE, GALEX and IUE archives. Considering the UV excess alone, \figref{fig:SED_comp} shows that it is challenging to detect the stripped star considered here if the companion is a O9V or B1V star. However, the UV excess can be used to search efficiently for hot star companions orbiting lower mass stars with later spectral types. For example, our stripped star heavily dominates in UV flux over the B8V companion shown in \figref{fig:SED_comp}. For comparison, the companion of the observed system HD~45166, which is the only clear case of an identified stripped star in this mass regime, is indeed a later type (B7V) star with a mass of about $5\, \Msun$ \citep{2005A&A...444..895S}\lang{; this is} consistent with the trend shown in \figref{fig:SED_comp}.

Another promising strategy is to search for the emission lines of stripped stars. Our models suggest that for certain combinations of stripped stars plus main sequence companions, these emission lines may be visible above the continuum of the companion. The most promising optical line is the HeII~$\lambda$4686 line. Further lines of interest are the  HeII~$\lambda$1640 line in the UV and a sequence of strong lines in the near IR, although the overall drop of intensity may make these more challenging to use. \lang{The} strength and shape of the emission \lang{lines depend} on the uncertain mass loss rates and terminal wind speeds. A search for these emission features \lang{recovers} only a subset of the population of stripped stars, with the mass ranges and orbital companions dictated by the actual values of the mass-loss rates and wind terminal speeds of stripped stars.

Further strategies include searches for radial velocity (RV) variations and  eclipses. \lang{These} RV searches are challenging since the expected variations for the brighter and more massive companion star are typically small and its lines are likely broadened by rotation \citep[][]{2014ApJ...782....7D}. \lang{The} RV variations in the possible emission lines of the companion would be larger, but they would need to be detected first. Binary systems hosting a stripped star that shows eclipses should be rare \lang{because of} the small radii of stripped stars ($\sim 1 \Rsun$). Moreover, the orbit is expected to have  widened as a result of previous mass transfer. However, they should appear in sufficiently large optical transient surveys, such as the Optical Gravitational Lensing Experiment \citep[OGLE,][]{2004AcA....54....1W, 2016arXiv161206394P}. \lang{Multicolor} eclipse data should allow identification of eclipses caused by a hot companion.

Finally, the strong EUV radiation of stripped stars could potentially be observed indirectly through emission lines in nearby gas that requires ionization by high-energy photons. The characteristic emission lines provide a diagnostic of the hardness of the ionizing source, which should in principle allow to differentiate between the presence of a stripped star or other ionizing sources such as a (accreting) neutron star or black hole. Similarly, the stripped stars may have potentially observable effects through irradiation of their companion. For example, they may induce differences between the day and night side of companion or they may partially ionize the disk of their companion, if present. Further simulations of the expected consequences are warranted.

\subsection{Failure to remove the H-rich envelope at low Z (I) -- implications for the ratio of SN type Ibc, IIb and II}

One of our key findings is that the Roche-lobe stripping process is inefficient at low metallicity. A thick layer of hydrogen-rich material is left at the surface of the stripped star when it contracts within its Roche lobe and mass transfer seizes. As discussed in \secref{sec:evol_Z_rlof}, we believe this is mainly the result of the lower opacity in the layers below the surface in these models. 

This has potentially important implications for the final surface composition of supernova progenitors \citep[cf. ][]{2017ApJ...840...10Y}. Roche-lobe stripping is considered as the main progenitor channel leading to type Ibc supernova, i.e.\ core collapse supernova that do not show signatures of H in their spectra \citep[e.g.][]{2007ApJ...670..747K,2010ApJ...725..940Y, 2011MNRAS.412.1522S, 2011A&A...528A.131C, 2013MNRAS.436..774E, 2016ApJ...818...75V, 2016MNRAS.461L.117E}.  At high metallicity Roche-lobe stripping removes effectively the entire H-rich envelope, for the orbital periods that we considered here. The small amount that remains is subsequently removed by the stellar wind before the stars \lang{explode}. We thus expect higher metallicity stripped stars to end their lives as a Ibc supernovae, \lang{which is} consistent with earlier \lang{works} \citep[cf.][]{2010ApJ...725..940Y, 2013MNRAS.436..774E}.  

However, we find very different results for low metallicity. For our \lang{metal-poor} models we find that the first Roche-lobe overflow event fails to remove the last $\sim 1\Msun$ of the envelope, which contains a mixture of hydrogen and helium. The total mass of pure hydrogen is $\sim 0.3\Msun$ at this stage. How much hydrogen is left at the final explosion, depends on whether or not the system evolves through a second mass transfer event during the helium shell burning phase. This depends on the initial orbital period and on the mass of the stripped star. \lang{We adopted} a relatively short orbital period in our simulations and we find that most, but not all,  of the hydrogen is removed during a second mass transfer. The final remaining hydrogen masses in our models are given in \tabref{tab:hestar_prop}. Our most metal-poor model evolves through a second mass transfer phase and has $M_{\text{H,tot}} = 0.03 \Msun$ left at the moment of explosion. This amount of hydrogen is small, but could be detected in early-time spectra \citep{2011MNRAS.414.2985D, 2012MNRAS.422...70H}, thus suggesting that the supernova could be classified as type IIb, if detected early enough. We refer to \citet{2017ApJ...840...10Y} for a discussion of the effect of mass and period. 

The general prediction from our models is thus that the fraction of type Ibc decreases for lower metallicity, while the fraction of type IIb rises, a conclusion that is also drawn by \citet{2017ApJ...840...10Y}. 

Determining these fractions observationally has proven to be challenging, since most surveys are biased to more massive galaxies, introducing a bias \lang{toward} metal-rich stellar populations \citep{2016ApJ...827...90L}. The most careful and comprehensive present-day analysis appears to be the work by \cite{2016arXiv160902921G, 2016arXiv160902923G}. In the most recent observational results a reduction of type Ibc rate is found for less massive galaxies. Given the general galaxy mass-metallicity relation this may indicate that type Ibc are less prevalent at lower metallicity \citep{2011ApJ...731L...4M, 2012IAUS..279...34A}, consistent with what our models predict. The type IIb rate appears to be unaffected. \lang{This} conclusion differs from the findings by earlier investigations, which showed a increase of the type IIb rate in dwarf galaxies \citep{2010ApJ...721..777A, 2012ApJ...759..107K, 2017ApJ...840...10Y}.

\subsection{Failure to remove the H-rich envelope at low Z (II) -- Implications for rapid population synthesis simulations including gravitational wave predictions}

Our results also have possibly important implications for the variety of rapid population synthesis simulations that rely on the fast but approximate algorithms by \cite{2000MNRAS.315..543H, 2002MNRAS.329..897H}. In these simulations stripped stars are approximated using evolutionary simulations for pure helium stars \lang{that were} computed by \citet{1998MNRAS.298..525P}. Our simulations show that this approximation is fairly accurate for high metallicity, but it breaks down at low metallicity.    

These algorithms form the basis of a large number of simulations that are used for a variety of predictions. This include results obtained with the StarTrack code \citep[e.g.][]{2010ApJ...715L.138B, 2012ApJ...759...52D, 2015ApJ...814...58D, 2016Natur.534..512B},  binary\_c \citep[][]{2004MNRAS.350..407I, 2006A&A...460..565I, 2009A&A...508.1359I, 2013ApJ...764..166D, 2014A&A...563A..83C, 2015ApJ...805...20S}\lang{,} and COMPAS \citep{2017NatCo...814906S}. 

The remaining hydrogen-rich envelope layer can potentially affect the further evolution of metal-poor binary systems. Further simulations will be needed to investigate this in more detail, but we can expect that a larger fraction of systems experience a second mass transfer stage when the stripped stars swells up during helium shell burning. The second phase of mass transfer occurs in our models when $Z \leq 0.0047$, \lang{but this} will depend on the adopted mass and orbital period. This may potentially be of interest in channels where the companion is an accreting white dwarf \citep{2002MNRAS.331.1027D, 2015MNRAS.451.2123T}. It could in principle affect the supernova type Ia rates through the single degenerate formation channel \citep{1973ApJ...186.1007W, 1982ApJ...253..798N}. If such systems enter a common envelope stage they may produce very tight binary systems that are of interest as gravitational wave sources. For gravitational wave sources in particular, current simulations predict the majority of sources to arise from metal-poor stellar populations, where we expect the effects to be largest.

\section{Summary \lang{and conclusions}}\label{sec:conclusions}

\lang{We investigated} the effect of metallicity on stars that lose their hydrogen-rich envelope through interaction with a companion. For this purpose we used the detailed stellar evolutionary code MESA and the atmosphere code CMFGEN.  Our findings apply to a typical massive binary (where the primary star has an initial mass of $12\Msun$) that fills its Roche lobe after leaving the main sequence but before the completion of helium burning  that avoids coalescence. We summarize our main results and their implications.

 \begin{enumerate}

 \item In agreement with earlier work, we find that Roche-lobe stripping exposes the helium core of the donor star and produces very hot and compact stars (80\,000\,K and $\sim 1 \Rsun$ in our solar metallicity reference model). \lang{These stars} fill the gap in mass between their higher mass counterparts, known as Wolf-Rayet stars, and their lower mass counterparts, subdwarf O and B stars. The stripped stars considered here are not expected as a result of single star evolution: they are uniquely produced as a result of binary interaction.
 
\item For single stars it is a \lang{well-known} fact that lowering the metallicity results in hotter and more compact stars, at least in the early evolutionary phases. Surprisingly, we find the opposite for stripped stars. At lower metallicity, mass donors shrink within their Roche lobe before the removal of the hydrogen-rich envelope is complete. We believe that this is due to the reduction of the opacity in the subsurface regions. The result is that metal-poor stripped stars are larger, more massive, more luminous, slightly cooler\lang{, and shorter lived} than their metal-rich counter parts.   
   
 \item Stripped stars are very efficient sources of ionizing photons. Despite losing about \lang{two-thirds} of their mass, the bolometric luminosities of stripped stars are comparable to their pre-interaction main sequence progenitors. However, most of the light is emitted in the extreme UV at wavelengths that are inaccessible by \lang{ground- and space-based} telescopes. Our reference model emits 85\% of its luminosity at wavelengths \lang{blueward} of the Lyman limit for H ionization and 60\% \lang{blueward} of the threshold to singly ionize helium. The flux at shorter wavelength is very sensitive to uncertainties in the mass loss rate. A corresponding single star of the same initial mass does not emit any significant number of ionizing photons during its helium burning phase. The number of ionizing photons is mildly dependent on metallicity. The HI and HeI ionizing photons vary by less than a factor of two among our models. 

 \item Stripped stars are not as luminous as massive Wolf-Rayet stars, which are emitting ionizing photons at the fastest rates in stellar populations. However, using a simple estimate we argue that they produce a comparable amount of ionizing photons. Stripped stars are favored by several effects. (a) They are the product of lower mass stars, which are favored by the initial mass function. (b) Their lower mass loss rates, give them relatively transparent atmospheres with a photosphere that lies close to their very hot stellar surfaces. This allows them to produce very hard radiation. (c) They evolve more slowly than their higher mass counterparts and spend a longer time in the phase during which they produce ionizing photons (about 1 Myr for the models considered here). (d) The ionizing photons are emitted with a long time delay after starburst, in contrast to those emitted by the short-lived massive Wolf-Rayet stars. This is interesting since it allows time for feedback from massive stars to remove most of the surrounding gas of the birth clouds that could trap the ionizing photons. We may thus expect a larger fraction of ionizing photons to escape and become available to ionize the intergalactic medium. A full assessment will require larger model grids. 
   
\item Our models predict that \lang{high metallicity} stripped stars have strong emission lines, a prediction that is robust against variations of a factor of three in the adopted wind mass loss rate. The strongest optical spectral feature of stripped stars is the HeII~$\lambda$4686 emission line. The HeII~$\lambda$1640 \lang{and H$\alpha$} show similar \lang{strengths}. Other characteristic features are the CIV~$\lambda$1548 and 1551 doublet and numerous emission lines of NIV and NV. These lines provide a promising way to identify stripped stars in the vicinity of their optically bright companion stars. 

 \item Our finding that Roche-lobe stripping fails to remove the complete H-rich envelope of metal-poor stars has implications for the further sequence of interaction phases that a binary system may undergo. Our results are in stark contrast with the approximate treatment of stripped stars in \lang{widely used} rapid binary population synthesis algorithms, where stripped stars are treated as pure helium stars. This is a fair approximation at high metallicity, but it breaks down at lower metallicity. This is a concern, especially for simulations for gravitational wave sources. These predict a dominating contribution from low metallicity, where we find the largest discrepancy. Follow-up studies of the implications are warranted. 
  
 \item Our results also have implications for the diversity supernova subtypes and in particular whether hydrogen is expected to be present in the final spectra. At high metallicity the combined effects of Roche-lobe stripping and winds remove the remaining hydrogen, as already pointed out in earlier work. At low metallicity, this is not true. After completing central helium burning, our lower metallicity models expand and fill their Roche lobe a second time. Traces of hydrogen are expected to be visible in the spectra of these supernova when they explode. This is consistent with the observationally derived decrease of type Ibc supernova in lower mass galaxies.   
 
 \end{enumerate}
 
\noindent  We conclude that advancing our understanding of stripped stars is of wide interest because of the many implications they have for astrophysics, ranging from questions about ionizing photons to formation scenarios for gravitational sources. It is humbling to realize that it has now been more than half a century since the first numerical simulations of this type were performed.  At this moment we have the luxury of \lang{an} improved understanding of the microphysics \lang{and increased} computational resources. For this study we \lang{explicitly chose} to limit the extent of the \lang{parameter} space, not because of the limitations in computational power, but because of the richness of processes that required discussion and deeper understanding.  However, to move forward from here, prioritization of efforts on the computational and observational side are needed. This will allow us to explore the physical processes and how they behave in the large parameter space and to increase the observed sample that can be used to verify and test the models.  
    
%


\begin{acknowledgements}

YG acknowledges first and foremost Emmanouil Zapartas and Mathieu Renzo for the stimulating daily interaction and discussions that were essential  for physical understanding, Pablo Marchant for his input concerning MESA and for making his scripts for Kippenhahn diagrams available\lang{,}  Mar\'ia Claudia Ram\'irez Tannus and Frank Tramper for their expertise in spectral models, Martin Heemskerk for providing computing expertise and support throughout the project and Alessandro Patruno for allowing us to use the Taurus computer.  
The authors further acknowledges Colin Norman, Hugues Sana, Norbert Langer, Tomer Shenar, Paul Crowther, Nathan Smith, Douglas Gies, Abel Schootemeijer, Miriam Garcia,  Onno Pols, Sung-Chul Yoon, JJ Eldridge, Elizabeth Stanway and Ed van den Heuvel for many stimulating discussions that have helped to shape this paper. We further acknowledge the referee for her/his time and helpful comments.
YG acknowledges Geneva Observatory for the hospitality and inspiring scientific environment during a collaboration visit and to Anna Watts for financial support of this visit through her Aspasia grant. Part of the simulations were conducted on the LISA computing cluster at surfSARA.  SdM acknowledges support by a Marie Sklodowska-Curie Action (H2020 MSCA-IF-2014, project id 661502). 
This research was supported in part by the National Science Foundation under Grant No. NSF PHY-1125915. 

\end{acknowledgements}

\bibliographystyle{aa}
\bibliography{references_bin}  
\appendix

\section{Connection between the atmosphere and structure models for varying metallicity}\label{app:stitch}

\begin{figure}
\centering
\includegraphics[width=0.8\hsize]{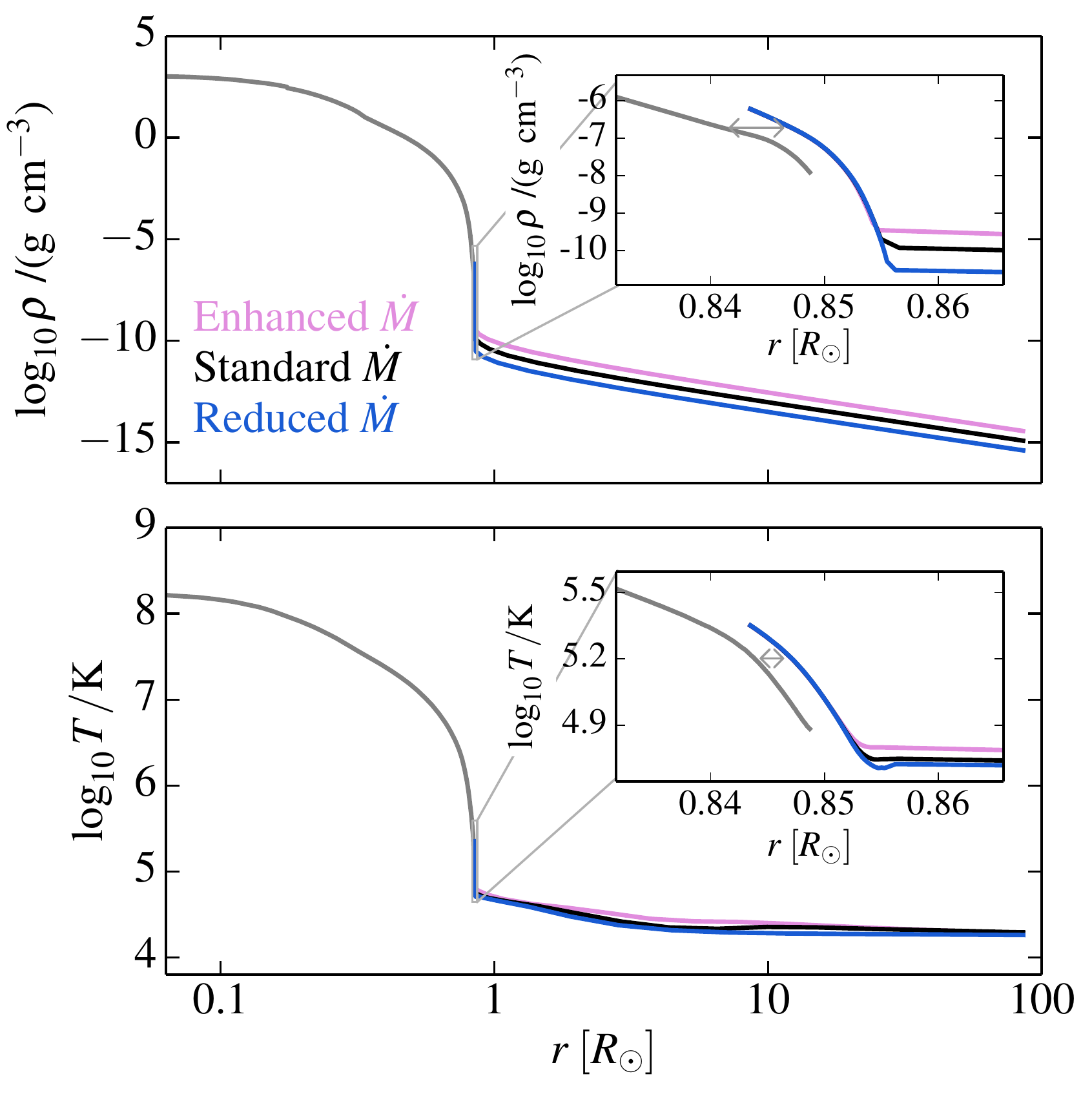}
\includegraphics[width=0.8\hsize]{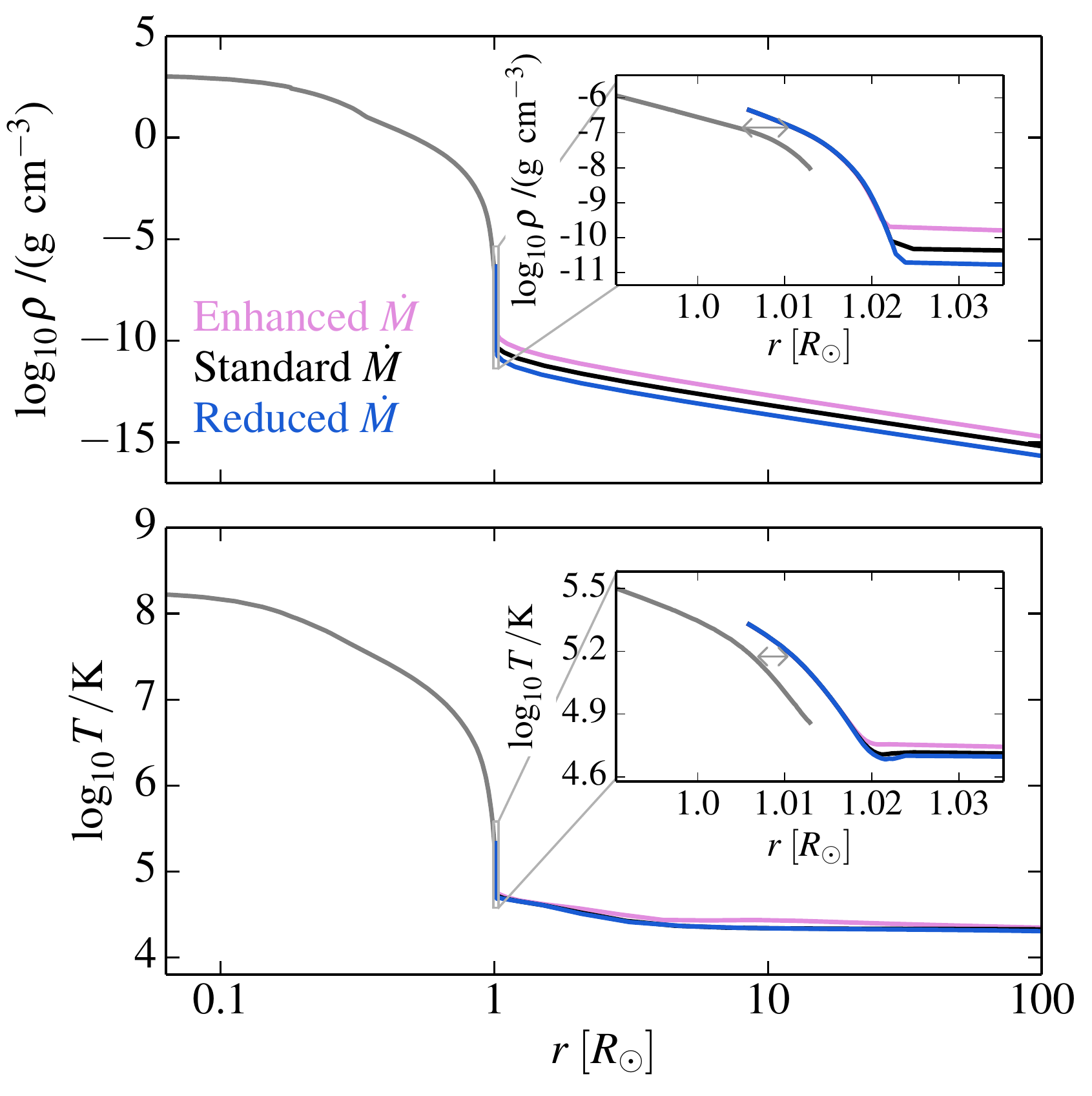}
\includegraphics[width=0.8\hsize]{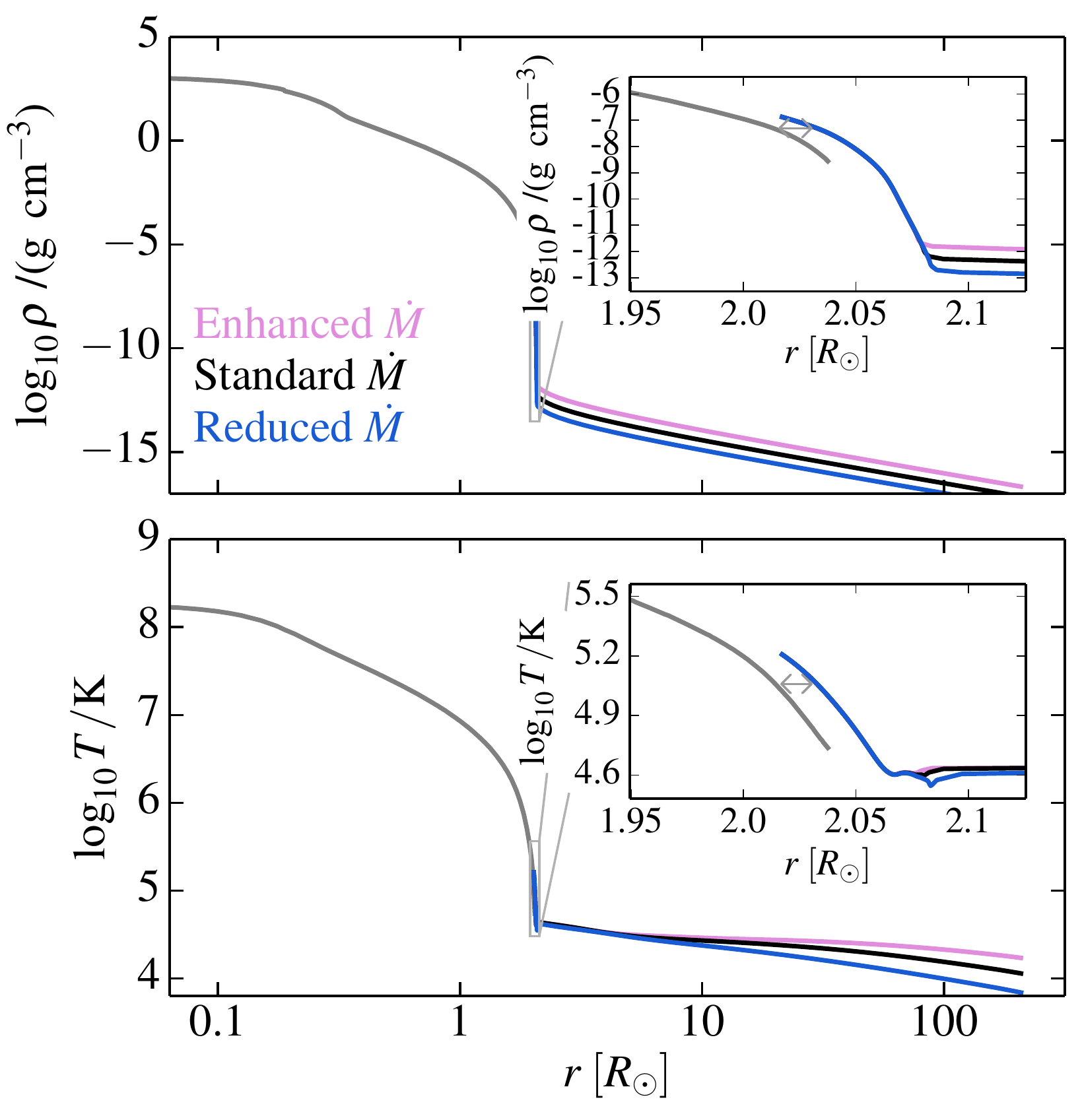}
\caption{\lang{Connection} between the MESA and CMFGEN models similar to \figref{fig:stitch_M12}, but here for metallicities $Z = 0.0047$ (top panel), $Z = 0.0021$ (middle panel) and $Z = 0.0002$ (bottom panel).}
\label{fig:stitch_Z}
\end{figure}

In \secref{sec:tailor} we describe how we construct CMFGEN atmosphere models for the stellar structure models computed with MESA. Here, we provide additional plots \lang{analogous} to \figref{fig:stitch_M12} to show the connection for the temperature and density structure. The top, middle\lang{,} and bottom \lang{panels} of \figref{fig:stitch_Z}, show our $Z = 0.0047$, $Z = 0.0021$\lang{,} and $Z = 0.0002$ models\lang{,} respectively. Within each panel\lang{,} we show the connection for the standard mass loss \lang{rate together} with a \lang{mass loss rate that is} three times enhanced and three times reduced.  
The differences in the estimated stellar radii from the MESA and CMFGEN models translate into temperature differences. The largest difference is $\sim 200$~K for the $Z = 0.0002$ model, which is negligible compared to the surface temperature ($\sim$50~000~K). This is accurate enough to reliably predict the spectral properties.

\section{Impact of parameter variations}\label{app:vinf}

\subsection{Wind speed and clumping variations}

\begin{figure*}
\centering
\includegraphics[width=\textwidth]{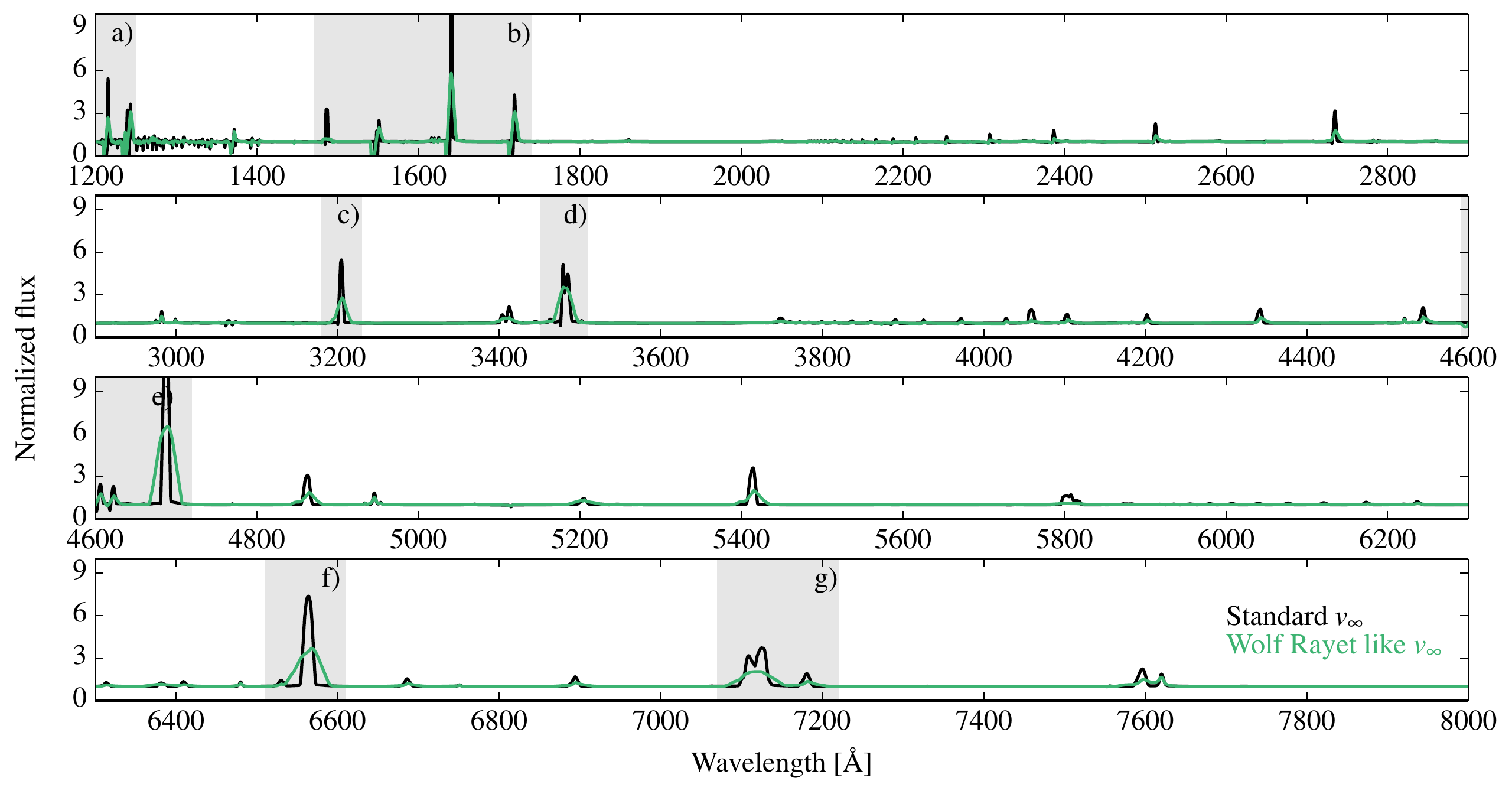}
\includegraphics[width=\textwidth]{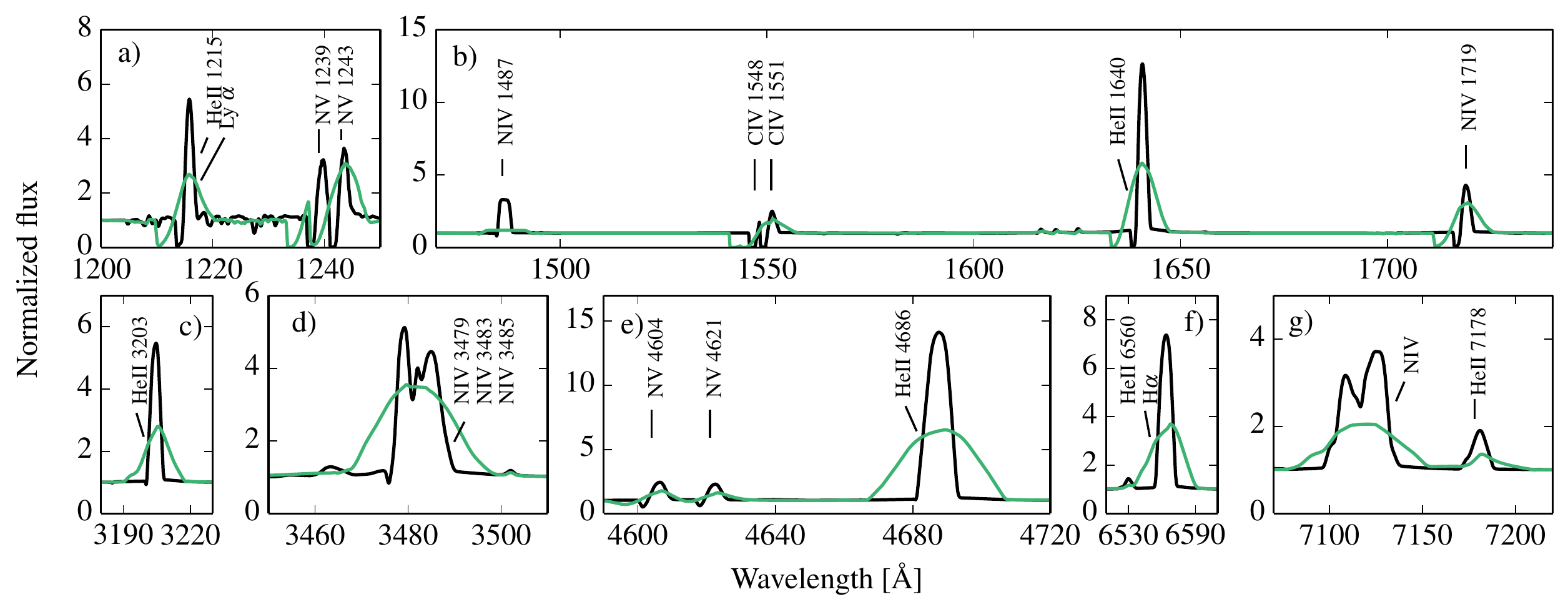} 
\caption{Illustration of the effect of the adopted terminal wind speed and clumping factor on the UV and optical spectrum. We show our reference model in black \citep[$v_{\infty} = 357$\kms and a volume filling factor of 0.5, based on][]{2008A&A...485..245G} and compare to a model, for which we adopt values that are typical for WR stars shown in green ($v_{\infty} = 1300$\kms and volume filling factor 0.1).}
\label{fig:spectrum_vinf}
\end{figure*}

For the atmosphere models presented in this work we \lang{adopted} a terminal wind velocity $v_{\infty} = 357$\kms motivated by the measurements for the stripped star in the HD~45166 system. This value is small in comparison with the typical values measured for the higher mass counterparts, WR stars, which commonly have $v_{\infty} = 1\,300$\kms \citep[see][for a discussion]{2008A&A...485..245G}. We \lang{investigated} the impact of changes in this assumption by computing an atmosphere model equivalent to our standard model, but instead assuming $v_{\infty} = 1\,300$~K and an increased effect of wind clumping by setting the wind clumping volume filling factor to 0.1.

The spectral energy distribution is not affected by the changes in these assumptions. However, the spectral features broaden and stand out less strongly above the continuum, as can be seen in \figref{fig:spectrum_vinf}.    
This would make it harder to identify the features of stripped stars when they are accompanied by an optically bright main sequence companion. However, the prominent HeII~$\lambda$4686 feature shown in panel e) is still about seven times stronger than the continuum, despite spanning over about 30~\AA.  

The example shown considers a more \lang{WR-like} set of parameters compared to the standard model. It is \lang{also interesting to} consider spectral models with low wind speed and increased clumping or high wind speed and less clumping. In the case of low wind speed and increased clumping we expect the spectral features to be stronger compared to the standard model as the increased wind clumping makes the atmosphere seem more optically thick. The lines would however remain narrow. For a model with instead high wind speed but less clumping we expect broad features, but potentially more P Cygni profiles compared to the model with high wind speed and more clumping presented in this subsection. This because less clumping makes the wind seem optically thinner.

\subsection{Other uncertain parameters}

Here we list several other uncertain parameters, which have not been investigated in detail in this work. We explain how these could potentially impact the appearance of the spectra.
\begin{itemize}
\item Composition not scaled according to solar. The relative metal mass ratios may vary with metallicity, in particular over cosmic time. A larger amount of CNO for a specific iron abundance would increase the spectral features of CNO, but not affect the emitted ionizing flux significantly.
\item Initial helium mass fraction. The initial helium mass fraction might vary between environments of the same metallicity. Such a change would affect the compactness of the stars throughout the evolution and also \lang{affect} the size of the helium core after the main sequence. This has direct implications for the luminosity and thus ionizing flux of the stripped stars. 
\item Including other Fe-group elements in spectral modeling. To include more elements from the Fe-group could increase the line blanketing in the extreme UV and in such a way reduce the flux of ionizing photons. 
\end{itemize}

\end{document}